\documentclass[iop,apj,twocolumn]{aastex631}

\usepackage{booktabs}

\shorttitle{Properties of Peg~III \& Psc~II} 
\shortauthors{Richstein et al.}

\begin{document}

\title{Structural parameters and possible association of the Ultra-Faint Dwarfs
 Pegasus~III and Pisces~II from deep Hubble Space Telescope photometry}

\author{Hannah Richstein}
\affil{Department of Astronomy, University of Virginia, 530 McCormick Road, Charlottesville, VA 22904, USA} 

\author{Ekta Patel}
\affil{Department of Astronomy, University of California, Berkeley, 501 Campbell Hall, Berkeley, CA 94720, USA} 
\affil{Miller Institute for Basic Research in Science, 468 Donner Lab, Berkeley, CA 94720, USA} 
\author{Nitya Kallivayalil}
\affil{Department of Astronomy, University of Virginia, 530 McCormick Road, Charlottesville, VA 22904, USA} 
\author{Joshua D. Simon}
\affil{Observatories of the Carnegie Institution for Science, 813 Santa Barbara Street, Pasadena, CA 91101, USA} 
\author{Paul Zivick}
\affil{Mitchell Institute for Fundamental Physics and Astronomy, Department of Physics and Astronomy, Texas A \& M University, 578 University Drive, College Station, TX 77843, USA}
\author{Erik Tollerud}
\affil{Space Telescope Science Institute, 3700 San Martin Drive, Baltimore, MD 21218, USA}
\author{Tobias Fritz}
\affil{Department of Astronomy, University of Virginia, 530 McCormick Road, Charlottesville, VA 22904, USA} 
\author{Jack T. Warfield}
\affil{Department of Astronomy, University of Virginia, 530 McCormick Road, Charlottesville, VA 22904, USA} 
\author{Gurtina Besla}
\affil{Steward Observatory, University of Arizona, 933 North Cherry Avenue, Tucson, AZ 85721, USA}
\author{Roeland P. van der Marel}
\affil{Space Telescope Science Institute, 3700 San Martin Drive, Baltimore, MD 21218, USA}
\affil{Center for Astrophysical Sciences, The William H. Miller III Department of Physics \& Astronomy, Johns Hopkins University, 3400 N. Charles Street, Baltimore, MD 21218, USA} 
\author{Andrew Wetzel}
\affil{Department of Physics \& Astronomy, University of California, Davis, One Shields Avenue, Davis, CA 95616, USA}
\author{Yumi Choi}
\affil{Space Telescope Science Institute, 3700 San Martin Drive, Baltimore, MD 21218, USA}
\affil{Department of Astronomy, University of California, Berkeley, 501 Campbell Hall, Berkeley, CA 94720, USA} 
\author{Alis Deason}
\affil{Institute for Computational Cosmology, Department of Physics, University of Durham, South Road, Durham DH1 3LE, UK} 
\affil{Centre for Extragalactic Astronomy, Department of Physics, University of Durham, South Road, Durham DH1 3LE, UK} 
\author{Marla Geha}
\affil{Department of Astronomy, Yale University, 52 Hillhouse Avenue, New Haven, CT 06520, USA}
\author{Puragra Guhathakurta}
\affil{UCO/Lick Observatory, Department of Astronomy \& Astrophysics, University of California Santa Cruz, 1156 High Street, Santa Cruz, CA 95064, USA} 
\author{Myoungwon Jeon}
\affil{Department of Astronomy \& Space Science, Kyung Hee University, 1732 Deogyeong-daero, Yongin-si, Gyeonggi-do 17104, Korea} 
\author{Evan N. Kirby}
\affil{Department of Astronomy, California Institute of Technology, 1200 E. California Blvd., MC 249-17, Pasadena, CA 91125, USA}
\affil{Department of Physics, University of Notre Dame, 225 Nieuwland Science Hall, Notre Dame, IN 46556, USA}
\author{Mattia Libralato}
\affil{AURA for the European Space Agency (ESA), Space
Telescope Science Institute, 3700 San Martin Drive, Baltimore, MD 21218, USA}
\author{Elena Sacchi}
\affil{Leibniz-Institut für Astrophysik Potsdam, An der Sternwarte 16, 14482 Potsdam, Germany}
\affil{INAF--Osservatorio di Astrofisica e Scienza dello Spazio di Bologna, Via Gobetti 93/3, I-40129 Bologna, Italy} 
\author{Sangmo Tony Sohn}
\affil{Space Telescope Science Institute, 3700 San Martin Drive, Baltimore, MD 21218, USA}

\received{2022 April 4}
\revised{2022 May 9}
\accepted{2022 May 19}
\published{2022 July 15}

\begin{abstract}

We present deep \textit{Hubble Space Telescope (HST)} photometry of the ultra-faint dwarf (UFD) galaxies Pegasus~III (Peg~III) and Pisces~II (Psc~II), two of the most distant satellites in the halo of the Milky Way (MW). 
We measure the structure of both galaxies, derive mass-to-light ratios with newly determined absolute magnitudes, and compare our findings to expectations from UFD-mass simulations. For Peg~III, we find an elliptical half-light radius of $a_h{{=}}1\farcm88^{+0.42}_{-0.33}$ ($118^{+31}_{-30}$~pc) and $M_V{=}{-4.17}^{+0.19}_{-0.22}$; for Psc~II, we measure $a_h{=}1\farcm31^{+0.10}_{-0.09}$ ($69\pm8$~pc) and $M_V{=}{-4.28}^{+0.19}_{-0.16}$. 
We do not find any morphological features that indicate a significant interaction between the two has occurred, despite their close separation of only $\sim$40~kpc.
Using proper motions (PMs) from $\textit{Gaia}$ early Data Release 3, we investigate the possibility of any past association by integrating orbits for the two UFDs in a MW-only and a combined MW and Large Magellanic Cloud (LMC) potential. We find that including the gravitational influence of the LMC is crucial, even for these outer-halo satellites, and that a possible orbital history exists where Peg~III and Psc~II experienced a close ($\sim$10--20~kpc) passage about each other just over $\sim$1~Gyr ago, followed by a collective passage around the LMC ($\sim$30--60~kpc) just under $\sim$1~Gyr ago. Considering the large uncertainties on the PMs and the restrictive priors imposed to derive them, improved PM measurements for Peg~III and Psc~II will be necessary to clarify their relationship. This would add to the rare findings of confirmed pairs of satellites within the Local Group.

\end{abstract}


\section{Introduction} \label{sec:intro}

\begin{figure*}
    \centering
    \includegraphics[width=0.97\textwidth]{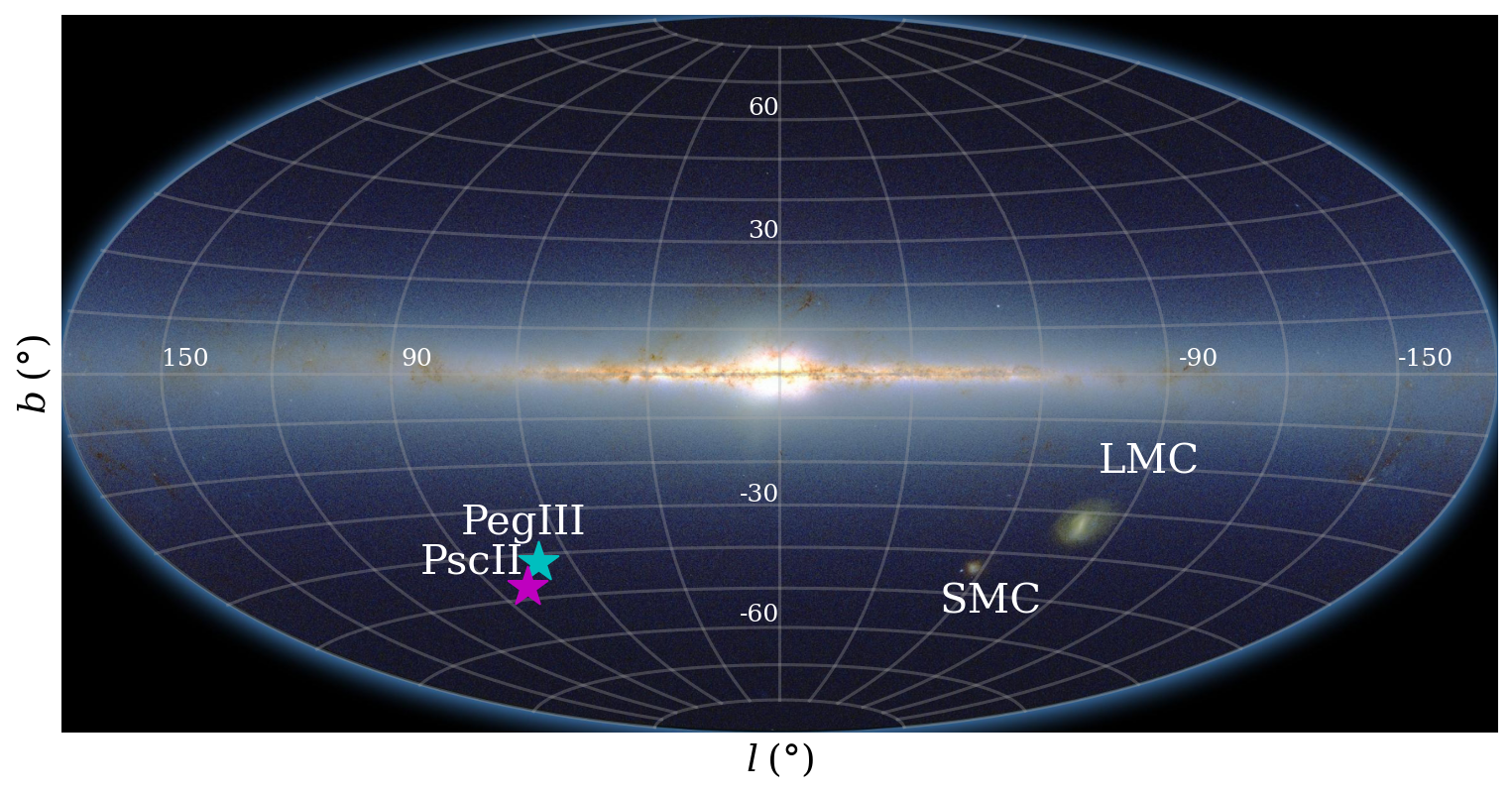}
    \caption{Peg~III (cyan star) and Psc~II (magenta star) shown relative to the Galactic plane, LMC, and SMC. Atlas Image mosaic obtained as part of the Two Micron All Sky Survey (2MASS), a joint project of the University of Massachusetts and the Infrared Processing and Analysis Center/California Institute of Technology, funded by the National Aeronautics and Space Administration and the National Science Foundation \citep{Skrutskie2006}.}
    \label{fig:proj}
\end{figure*}

Ultra-faint dwarf (UFD) galaxies are the most dark-matter-dominated systems discovered, and thus a preferred laboratory for studying how well cosmological models predict behavior on small scales. The widely accepted Lambda Cold Dark Matter ($\Lambda$CDM) model uses the hierarchical accretion of low-mass systems to explain the growth of dark matter halos \citep[e.g.,][]{NFW1997}. As UFDs occupy the least-massive dark matter halos discovered, they could be considered remnants of this hierarchical process while themselves having formed prior to the epoch of reionization and undergoing little evolution since then \citep[e.g.,][]{Ricotti2005,Gnedin2006,Bovill2009}.

The dark matter halos in which UFDs reside have extrapolated virial masses of approximately 10$^9$~M$_{\odot}$ \citep[e.g.,][]{Strigari2008}, about two orders of magnitude smaller than where the $\Lambda$CDM theory 
predicts central dark matter densities in apparent contrast with observations.
Dark-matter-only simulations predict cusps, while observations of dwarf spiral and dwarf spheroidal galaxies (dSphs) show cored mass distributions \citep[e.g.,][]{Flores1994,Moore1994,Read2005,Goerdt2006}. Other analyses have suggested that dSphs are consistent with the expected dark matter density profiles \citep[e.g.,][]{Jardel2013,Breddels2013,Strigari2017,Read2018,Read2019}.


A limiting factor for observational constraints is that we are often restricted to line-of-sight (LOS) velocity data of the stars residing in dwarf galaxies, and with no information on the tangential velocity components, we suffer from the mass-anisotropy degeneracy.
It will thus take more information, such as the shape of the velocity distribution or galactic internal proper motions (PMs), in addition to radial velocities to be able to distinguish a dark matter central core or cusp \citep[e.g.,][]{Strigari2007,Read2021,Guerra2022}.
Once we have full kinematic information, current virial mass estimators can be extended to further constrain UFD dark matter halo properties \citep{Errani2018}.
In the meantime, we can use more easily measured UFD properties such as half-light radius and luminosity to explore the population as a whole, as well as their simulated analogs.

Defined by having $M_V$ values fainter than $-7.7$ \citep[e.g.,][]{Bullock2017,Simon2019}, 
UFDs went undetected until the advent of large-scale digital sky surveys, beginning with the Sloan Digital Sky Survey in 2005 (SDSS; \citealt{York2000,Willman2005}). Currently, over 21 UFDs have been spectroscopically verified, and more than 20 other candidates have been identified \citep[e.g.,][]{Simon2019}. These numbers are expected to increase further with the beginning of data collection at the Vera C. Rubin Observatory.

While there is much to learn from studying any of these faint satellites, examining the relatively isolated UFDs at larger Galactocentric distances is particularly useful, as 
they are more comparable to those produced in most simulations that resolve down to the UFD-mass level ($M_{\mathrm{vir}} \simeq 10^9$~M$_{\odot}$ at $z{=}0$) \citep[e.g.,][]{Jeon2017,Jeon2021a,Jeon2021b,Wheeler2019}. 
More recently, \cite{Applebaum2021} and \cite{Grand2021} were also able to resolve down to UFD-mass scale using cosmological MW zoom-in simulations.  
Comparing the properties of observed UFDs to both of these types of simulations could help us to disentangle the effects that host galaxies may have on their satellite UFDs. 

Pegasus~III (Peg~III) is located at a heliocentric distance of approximately 215~kpc ($R_{\mathrm{GC}}{\sim}213$~kpc; \citealt{Kim2016}) and is thus one of the most distant MW UFDs known. 
Peg~III was reported in \cite{Kim2015}, having been discovered in Data Release 10 of SDSS \citep{Ahn2014} 
and confirmed with the Dark Energy Camera (DECam). 
The discovery team noted the proximity ($\sim$30~kpc projected and $\sim$32~kpc line-of-sight separation) of Peg~III to Pisces~II (Psc~II; $R_{\odot}{\sim}$183~kpc, $R_{\mathrm{GC}}{\sim}$182~kpc; \citealt{Belokurov2010,Sand2012}) at the time and suggested the possibility of an association. 
Figure \ref{fig:proj} shows where Peg~III and Psc~II lie in relation to the Galactic Plane, as well as the LMC and SMC \citep{Skrutskie2006}.
In a follow-up paper using Magellan/IMACS for photometry and Keck/DEIMOS for spectroscopy, \cite{Kim2016} derived a radial velocity for Peg~III that, in the Galactic standard-of-rest (GSR), only differed from that of Psc~II by $\sim$10~km~s$^{-1}$ ($-67.6\pm2.6$ and $-79.9\pm2.7$~km~s$^{-1}$ \citep{Kirby2015}, respectively), and calculated a 3D-separation of $\sim$43~kpc. Their team also found Peg~III to have an irregular shape elongated in the direction of Psc~II.

More recently, \cite{Garofalo2021} used the Large Binocular Telescope (LBT) to study variable stars in both UFDs. Using isodensity contour maps, they found no support for a physical connection between Peg~III and Psc~II, as neither UFD appeared to have an irregular shape. They suggested that the regular structures of both UFDs eliminate the notion of a stellar stream or another clear link between them.
However, even if the two UFDs themselves have not interacted, the possibility of them infalling as a pair or as part of a group is not precluded \citep[e.g.,][]{Wetzel15}.

Here, we present new, deep \textit{Hubble Space Telescope} (\textit{HST}) imaging of Peg~III and Psc~II, allowing further exploration of how these two UFDs may or may not be associated. We produce photometric catalogs and derive structural parameters, integrated $V$-band magnitudes, and mass-to-light ratios. We also conduct an orbital analysis of the two UFDs using proper motions (PMs) from \textit{Gaia} 
\citep{gaia1}
early Data Release 3 (\citealt{GaiaeDR3};\ eDR3). In Section \ref{sec:data}, we present our data and describe how they were processed. We measure the structural parameters and calculate mass-to-light ratios in Section \ref{sec:struc}. In Section \ref{sec:orbits}, we use $Gaia$ eDR3 to examine whether Peg~III and Psc~II could have had a past interaction. We discuss our results and conclude in Sections \ref{sec:disc} and \ref{sec:conc}, respectively.

\section{Hubble Space Telescope Data} \label{sec:data}
\subsection{Observations}
The observations of Peg~III and Psc~II were performed using the F606W and F814W filters of the \textit{HST} Advanced Camera for Surveys (ACS) Wide Field Channel (WFC) as part of Treasury program GO-14734 (PI: Kallivayalil). Parallel, off-target fields were simultaneously taken with the Wide Field Camera 3 (WFC3) using the same filters on the UV/visible (UVIS) channel for the purpose of learning more about the UFD stars farther from their galactic centers (when applicable) and being able to better characterize the stellar background distributions.
The Peg~III observations were taken on 2017 April 26 and 2017 May 2 using two orbits for F606W and two orbits for F814W. Psc~II was observed between 2017 June 19 and 2017 July 12 with two orbits dedicated to each filter. Each pair of ACS exposures totaled 4744~s, while each WFC3 pair totaled 5117~s. The long integration times 
allowed us to reach a signal-to-noise ratio (S/N) of 250 at $m_{\rm{F606W}}{=}23$.
The observations for both galaxies covered a single ACS $202'' \times 202\arcsec$ field and a single WFC3 $162'' \times 162\arcsec$ off-field. The visits for each UFD were performed within a restricted orientation range and over a four-point dither pattern, optimized for the astrometric goals of this Treasury program, but suitable for photometry as well.

\subsection{Reduction and Photometry}

\begin{figure*}
    \centering
    \includegraphics[width=0.97\textwidth]{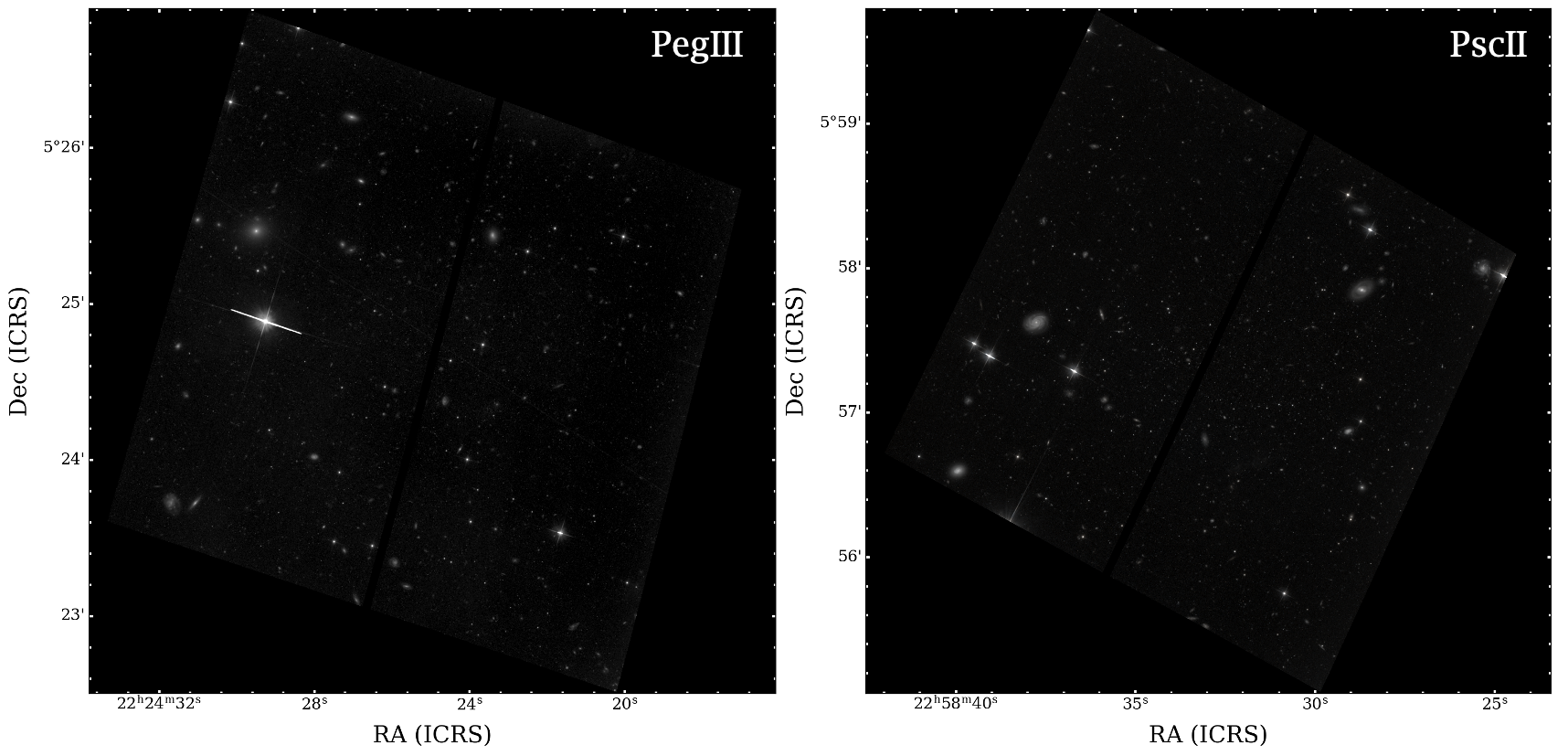}
    \caption{False-color images of the ACS fields for Peg III (left) and Psc II (right). The F606W image is used for the blue channel, the F814W for the red channel, and an average of the two for the green channel \citep{aplpy}. At the distance of Peg III \citep[215 kpc][]{Kim2016}, 1$\arcmin$ corresponds to $\sim$63 pc. For Psc II at 183 kpc \citep[][]{Sand2012}, 1$\arcmin$ corresponds to $\sim$53 pc.}
    \label{fig:rgb}
\end{figure*}

\begin{figure*}
    \centering
    \includegraphics[width=0.8\textwidth]{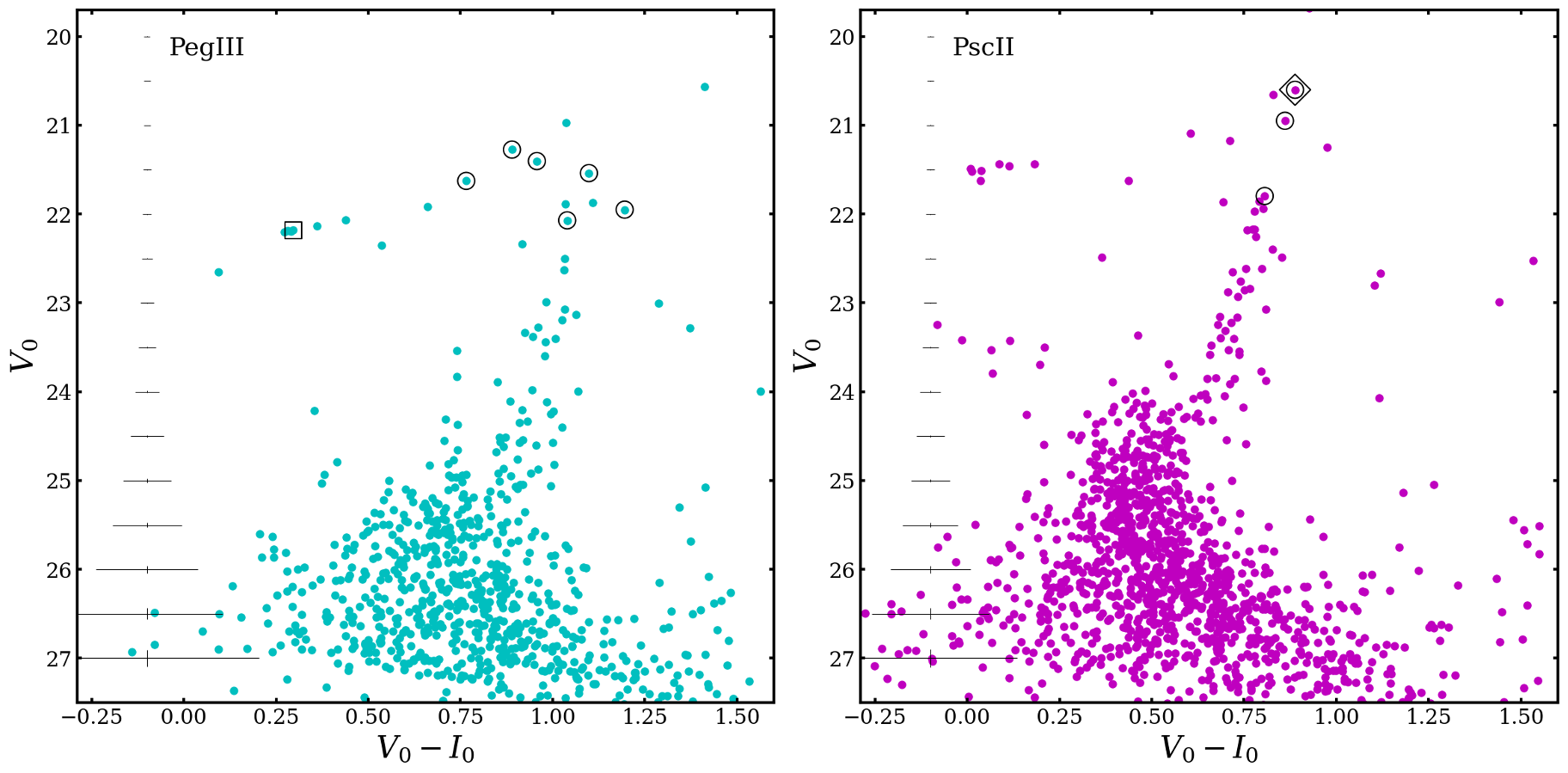}
    \caption{Color-magnitude diagrams of  Pegasus~III (left) and Pisces~II (right) in VEGAMAG. The typical color and magnitude errors are shown to the left. Confirmed spectroscopic members in our catalog are shown in circles for Peg~III, along with a star that has ambiguous membership in a square \citep{Kim2016}. There is one Peg~III spectroscopic member that is not included in our catalog due to it falling within the ACS chip gap. The Psc~II spectroscopic members with matches in our catalog are shown in circles \citep{Kirby2015}. Only 3 of the 7 member stars fall within the ACS field-of-view (FOV). The spectroscopic member with a \textit{Gaia} proper motion (PM) is in the diamond. The other stars used in the Psc~II PM measurement were either too bright or not in the ACS FOV. Only sources flagged as stars are shown.}
    \label{fig:cmd}
\end{figure*}

The images were processed and corrected for charge-transfer inefficiency (CTI) using the current ACS and WFC3 pipelines. In each filter, the four dithered images were combined using the DRIZZLE package \citep{Fructer2002} to create \texttt{drc} fits files. 
False-color images \citep{aplpy} of the two drizzled fields  are shown in Figure \ref{fig:rgb}.
Jackknife resampling was performed on the separate dither images to create four three-dither-combined images for deriving empirical errors.
We used the \texttt{photutils} \citep{Bradley2020} routines \texttt{DAOStarFinder} and \texttt{aperture\_photometry} to detect sources and calculate the flux inside two sizes of circular apertures. After analyzing flux counts from different apertures, we found that a four-pixel radius was both large enough to capture the concentrated stellar light and small enough avoid the inclusion of light from neighboring sources. By extending the aperture radius by two pixels to create a six-pixel radius, we could determine whether a source was more extended and thus not a star.

We used the four- and six-pixel instrumental magnitudes ($m_{\mathrm{inst}}$) to create a flag differentiating stars from galaxies. First, we calculated the median magnitude difference between the four- and six-pixel radius values and then
determined the uncertainty on the difference between the four- and six-pixel binned $m_{\mathrm{inst}}$ values as a function of the four-pixel $m_{\mathrm{inst}}$.  
After accounting for an error floor, we fit a linear relation to these uncertainties as a function of flux. 
We considered a source to be a star if its magnitude difference from the median was within two sigma of the fitted functional value. 
This criterion excluded galaxies, which showed larger magnitude differences as well as deviation from the fitted relation between the four- and six-pixel $m_{\mathrm{inst}}$.
After applying the flags, we accounted for the encircled energy corrections on the four-pixel radius flux values, converted the flux to STMAG, and matched sources between the F606W and F814W images. If a source was flagged as a star in either filter, it was used in the analysis for this paper.

Sources in the three-dither combined images went through the same aperture photometry pipeline and were matched across the four combinations in each filter using a 6-parameter linear transformation. To derive empirical errors, we took the standard deviation of the magnitudes of the sources found across all of the combined images. 
The sources were matched in the F606W and F814W filters, then matched to the \texttt{drc} source list using the same 6-parameter transformation. 

The observed magnitudes were corrected for dust extinction and reddening using the \texttt{dustmaps} module \citep{Green2018} with the \cite{Schlegel1998} maps and \cite{Schlafly2011} recalibration.
We converted the ACS photometry from STMAG filters to VEGAMAG $V$ and $I$ using the conversions and zero-points given in \cite{Sirianni2005}. To convert the WFC3 photometry, we employed \texttt{synphot} \citep{synphot} to generate representative blackbodies of field stars (5200~K and 5100~K for the Peg~III and Psc~II off-fields). Finally, we used \texttt{stsynphot} \citep{stsynphot} to calculate the filter conversion terms for these blackbodies and the WFC3 VEGAMAG zero-points for the observation dates. The resulting color-magnitude diagrams (CMDs) and typical errors are shown in Figure \ref{fig:cmd}, where cuts have been made based on the ``star-galaxy" flag. 

\begin{figure*}
    \centering
    \includegraphics[width=0.8\textwidth]{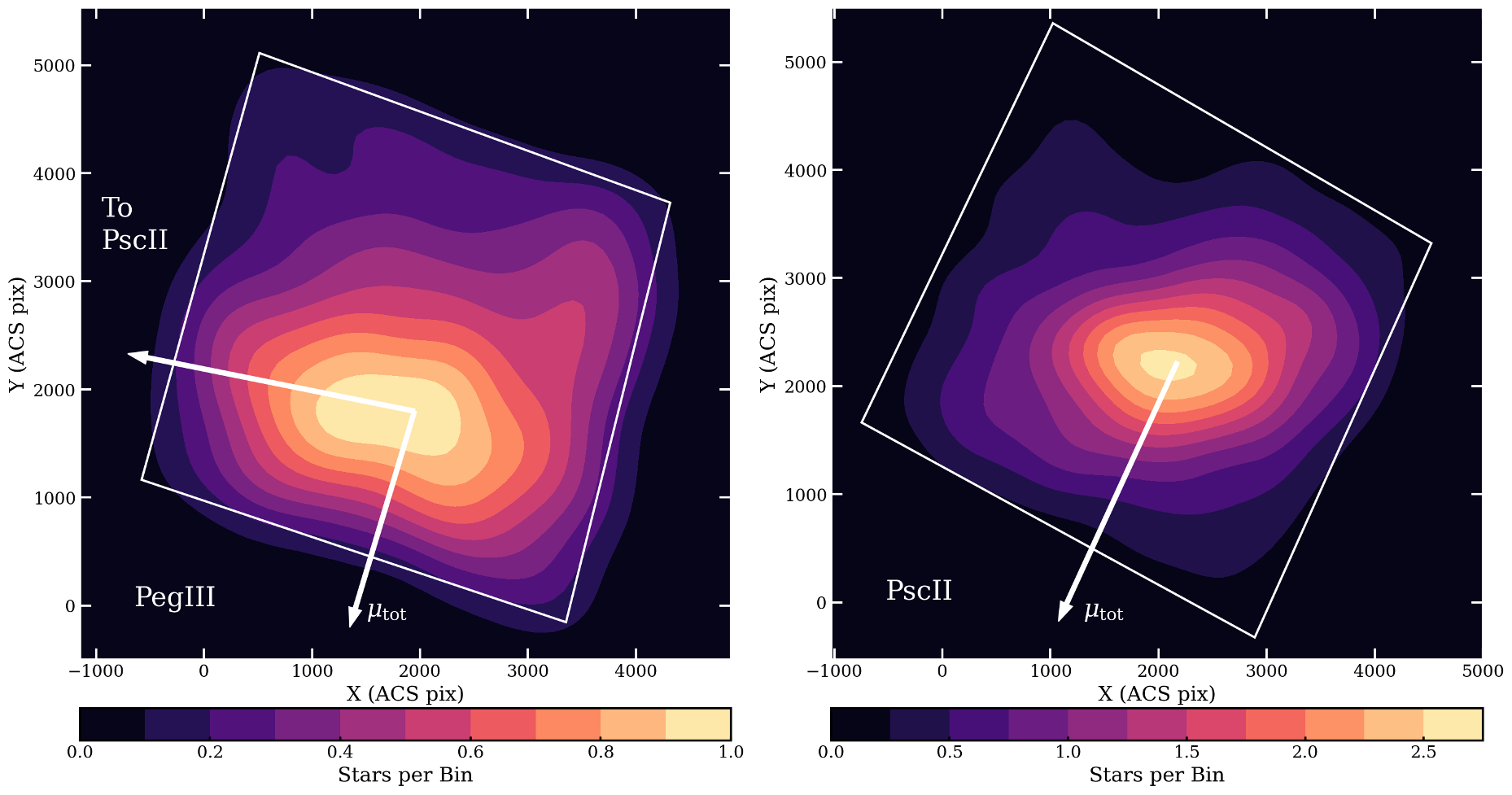}
    \caption{Density contour maps of Peg~III (left) and Psc~II (right). Each map shows 10 levels defined by the color bar and in units of stars per $4\farcs5\times4\farcs5$ bin. Note that the two color bars have different ranges and are not integer values as they are from kernel density estimates. The white lines show the borders of the ACS FOV, and the white arrows represent $\mu_{\mathrm{tot}}$ over a timespan of 0.5~Myr from \cite{mcconnachie21b}. In this and all subsequent plots, the galaxies are oriented such that North is in the direction of the increasing $y$-axis and East is in the direction of the decreasing $x$-axis. The white arrow extending from the center of Peg~III to the East is pointing in the direction of Psc~II on the sky.} 
    \label{fig:contours}
\end{figure*}

To present the data and to explore the existence of the elongation in Peg~III measured by \cite{Kim2016}, we created a contour map of stellar number density by performing a kernel density estimate on the data using 67 4\farcs5 by 4\farcs5 bins. 
The maps are orientated such that north is in the direction of the positive $y$-axis and east is in the direction of the negative $x$-axis.
The left panel of Figure \ref{fig:contours} shows the 10 levels as filled contours for Peg~III. 
There is an overdensity in the northwest (NW), however it is not in the direction of Psc~II, which is indicated by the white arrow pointing from the center to the east.
The contour map for Psc~II is illustrated in the right panel of Figure \ref{fig:contours}, created using the same process as for the Peg~III map. This is consistent with density contour maps from \cite{Belokurov2010}, \cite{Sand2012}, \cite{Munoz2018}, and \cite{Garofalo2021}, as all show Psc~II with more regular levels. Comparing Psc~II to Peg~III, we see that Psc~II is more compact and has a higher peak smoothed surface density.

\section{Structural Analysis} \label{sec:struc}
\begin{figure*}
    \centering
    \includegraphics[width=0.7\textwidth]{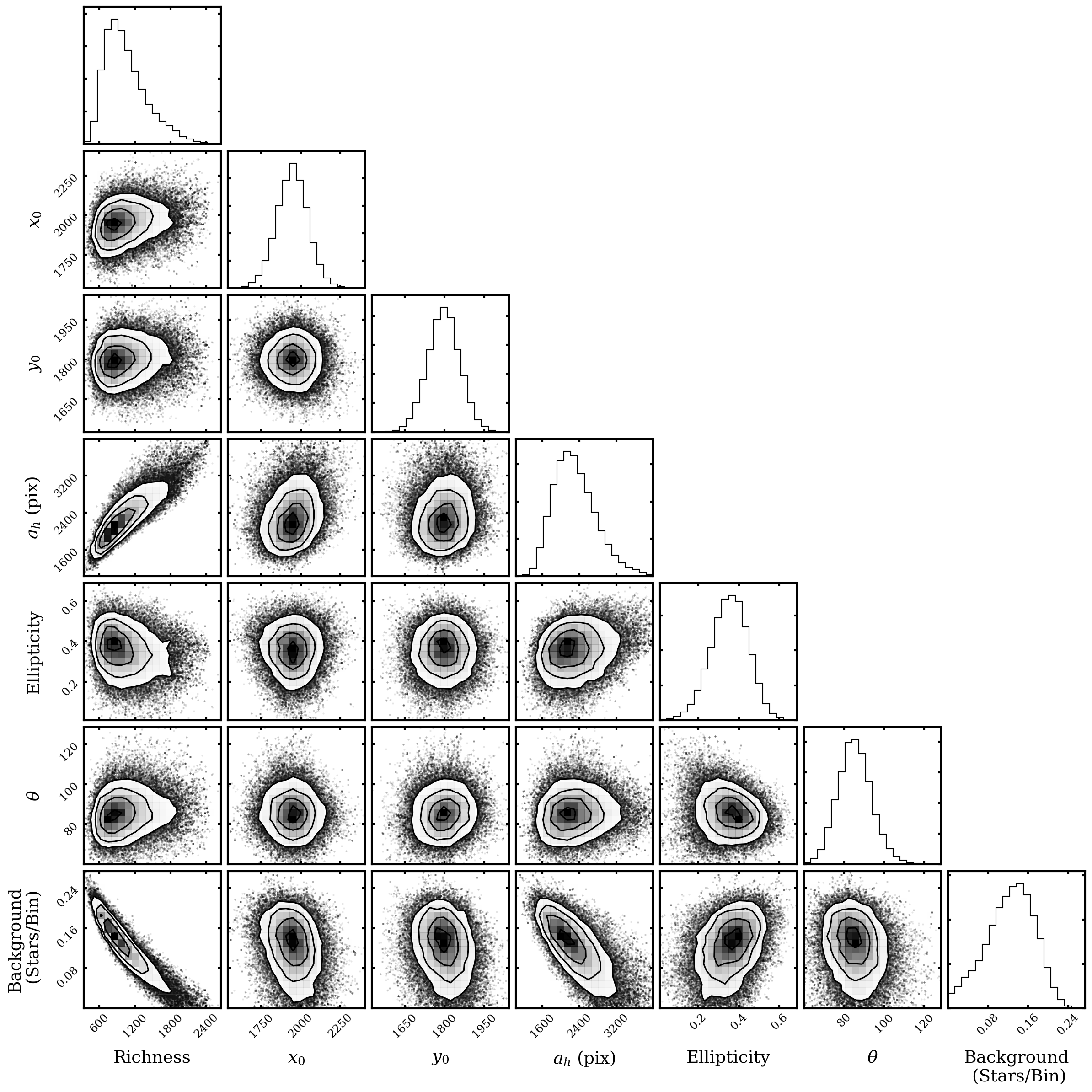}
    \caption{Corner plot for the  Pegasus~III structural parameters from the exponential fit. Positive correlations exist between the richness (number of stars) and elliptical half-light radius as well as the elliptical half-light radius and ellipticity. There is a negative correlation between the richness and background density.}
    \label{fig:cornerPegExp}
\end{figure*}
\subsection{Fitting 2D-Profiles}

To model the spatial structure of Peg~III and Psc~II, we followed the technique described in \cite{DrlicaWagner2020} and \cite{Simon2021}, largely based on the method shown in \cite{Martin2008}. We modeled each UFD with exponential and Plummer (\citeyear{Plummer2011}) profiles and performed binned Poisson maximum likelihood fits to the probability density functions with the following free parameters: center position ($x_0,y_0$), richness (number of stars), 2D, projected semimajor axis of the ellipse that contains half of the total integrated surface density of the galaxy (elliptical half-light radius; $a_h$), ellipticity ($\epsilon$), position angle of the semimajor axis measured from North through East ($\theta$), and background surface density (average density of stars in the field not belonging to the galaxy) ($\Sigma_b$). In past literature on these UFDs, the elliptical half-light radius $a_h$ has often been referred to as $r_h$, but here we have chosen to make the explicit distinction between $a_h$ and the azimuthally-averaged half-light radius $r_h$ (equal to $a_h\sqrt{1-\epsilon}$) for clarity and the purpose of our comparison to simulations.

The normalized functional forms of the exponential and Plummer profiles are as follows:
\begin{equation}
    \Sigma_{\mathrm{exp}}(r_i) = \frac{1}{2\pi r_e^2 (1-\epsilon)}\textnormal{exp}\bigg({-}\frac{r_i}{r_e}\bigg) 
    \label{eq:pdf_exp}
\end{equation}

\begin{equation}
    \Sigma_{p}(r_i) = \frac{r_p^2}{2\pi (1-\epsilon)} (r_i^2 + r_p^2)^{-2}.  
    \label{eq:pdf_pl}
\end{equation}

\noindent Here, $r_e$ and $r_p$ are the scale lengths for each respective model, with $r_e${=}$a_h/1.68$ and $r_p${=}$a_h$, and the first term in each equation is the normalization term, set to integrate to unity over all space.
The $r_i$ term is the elliptical radius of source $i$, defined as
\begin{equation}
    r_i=\bigg\{\bigg[ \frac{1}{1-\epsilon} (X_i\textnormal{cos}\theta-Y_i\textnormal{sin}\theta)\bigg]^2 + (X_i\textnormal{sin}\theta + Y_i\textnormal{cos}\theta)^2\bigg\}^{1/2}.
    \label{eq:r_ell}
\end{equation}
\noindent $X_i$ and $Y_i$ are the spatial offsets from the centroid, where $X_i=x_i-x_0$ and $Y_i=y_i-y_0$.

To calculate the best-fitting parameters, we used the Markov chain Monte Carlo (MCMC) ensemble sampler \texttt{emcee} \citep{ForemanMackey2013}. After adding the background surface density term, we fit the following two functions:
\begin{equation}
    \Sigma_{\mathrm{exp,tot}}(r_i) = \Sigma_{\mathrm{exp}}(r_i) + \Sigma_b
    \label{eq:exp_fit}
\end{equation}

\begin{equation}
    \Sigma_{p\mathrm{,tot}}(r_i) = \Sigma_{p}(r_i) + \Sigma_b .
    \label{eq:pl_fit}
\end{equation}

\begin{figure*}
    \centering
    \includegraphics[width=0.7\textwidth]{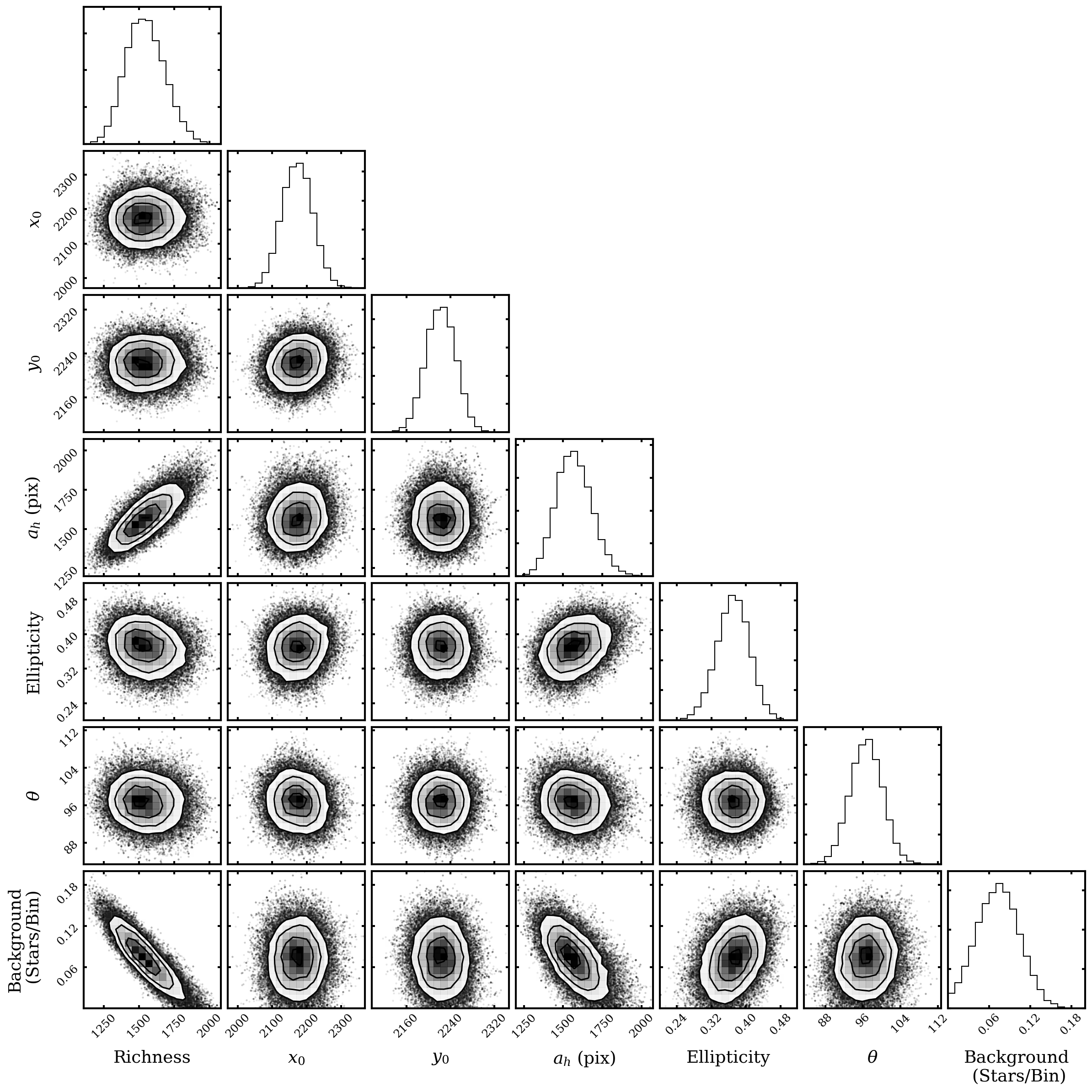}
    \caption{Same as Fig. \ref{fig:cornerPegExp}, for Pisces~II.}
    \label{fig:cornerPscExp}
\end{figure*}

\begin{figure*}
    \centering
    \includegraphics[width=\textwidth]{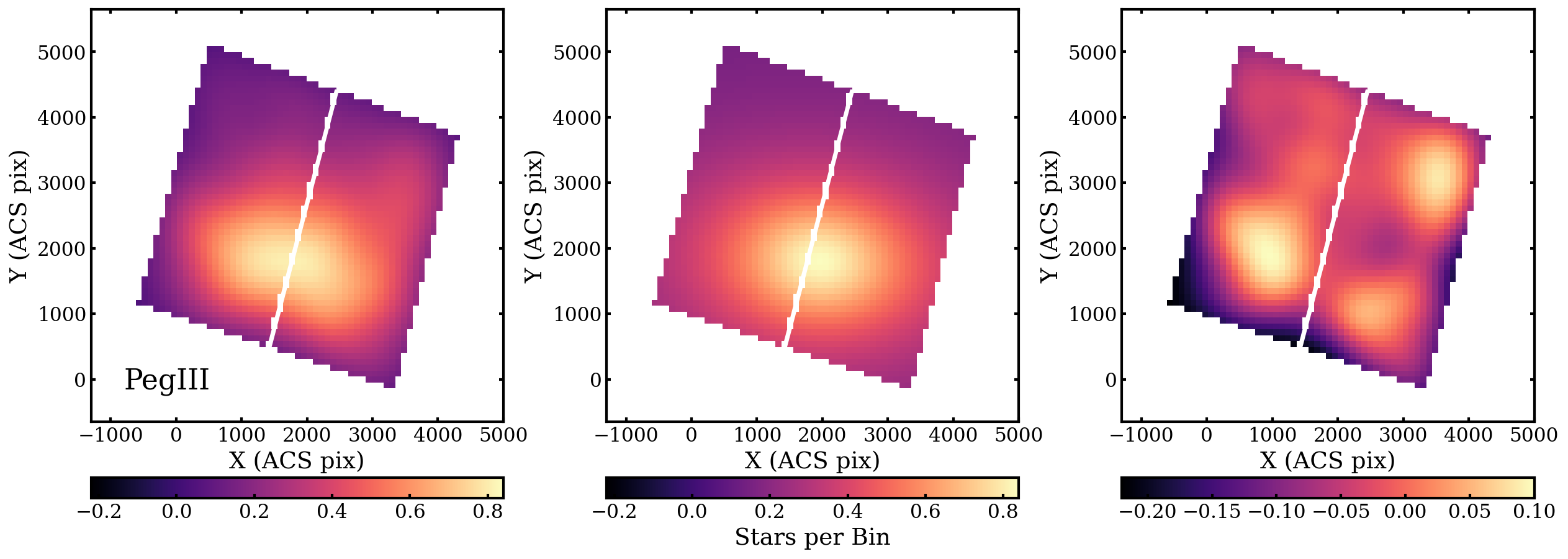}
    \caption{\textbf{Left}: Peg~III data, smoothed using a Gaussian kernel with a FWHM of 0\farcs5. \textbf{Middle}: Best-fitting exponential model, smoothed with a 0\farcs5 FWHM Gaussian kernel. \textbf{Right}: The residuals between the smoothed data and the model.}
    \label{fig:dmrPeg3}
\end{figure*}

\begin{figure*}
    \centering
    \includegraphics[width=\textwidth]{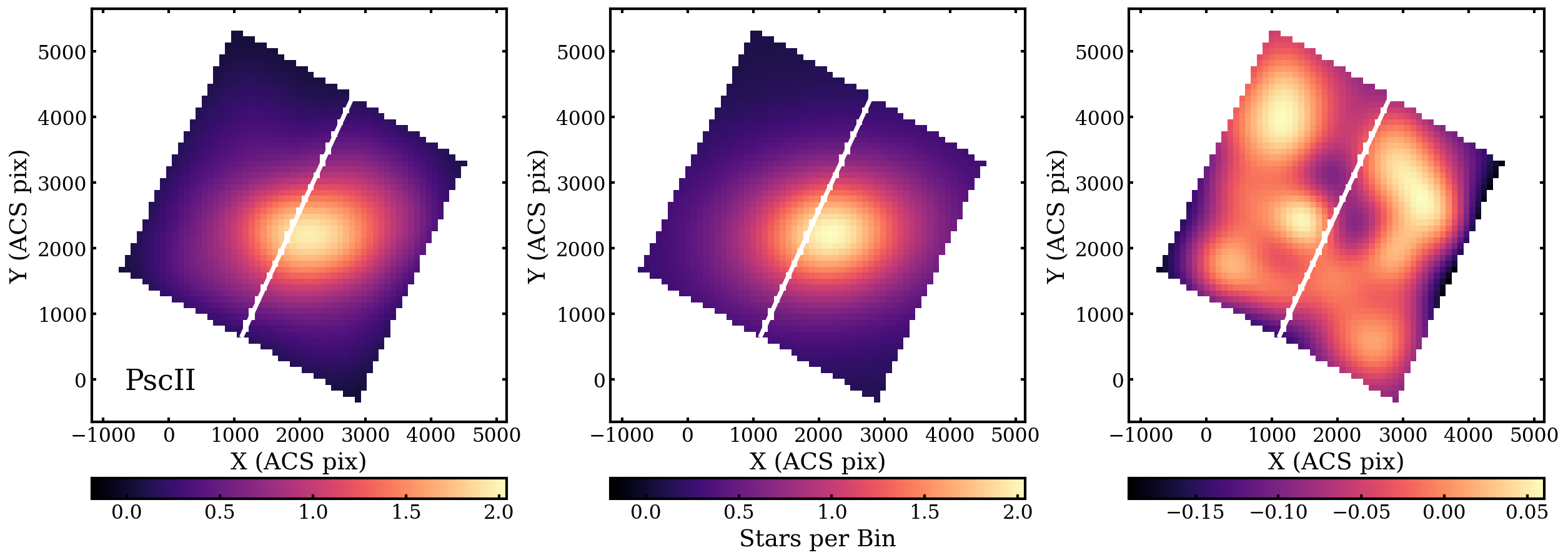}
    \caption{\textbf{Left}: Psc~II data, smoothed using a Gaussian kernel with a FWHM of 0\farcs5. \textbf{Middle}: Best-fitting exponential model, smoothed with a 0\farcs5 FWHM Gaussian kernel. \textbf{Right}: The residuals between the smoothed data and the model.}
    \label{fig:dmrPsc2}
\end{figure*}

We created $4\farcs5$ by $4\farcs5$ bins across the ACS field-of-view (FOV), masking the area of the chip gap and outside the FOV, and counted the number of stars in each bin.
We chose this bin size as it was large enough that the highest surface density areas of these sparse galaxies contained more than one star per bin and small enough that it would not obscure any structural features.
The corner plots \citep{corner} and posterior distributions for the exponential fits are shown in Figures \ref{fig:cornerPegExp} and \ref{fig:cornerPscExp} for Peg~III and Psc~II, respectively. For these and many subsequent figures, we choose to show plots using the exponential fit values as there are not considerable differences between the two profiles, and it allows easy comparison to what has already been presented in the literature. The values from this work and past works that measured structural parameters are shown in Tables \ref{tab:peg3} and \ref{tab:psc2}, with the center position transformed from pixel space to coordinates in right ascension (RA) and declination (DEC). The spatially-binned, smoothed data, the smoothed best-fit exponential model, and the residuals are shown in Figures \ref{fig:dmrPeg3} and \ref{fig:dmrPsc2}. 
For Peg~III, the two most prominent residuals are in the NW and southeast (SE), with another to the southwest (SW). These show that a pure elliptical model cannot describe the full morphological complexity of Peg~III.
Psc~II has prominent residuals to the north and NW, as well as one near the center of the image. We expect such small residual variation to be present even for models that reproduce the galaxy morphology.

\begin{table*}
\centering
\begin{tabular}{llll}
\multicolumn{4}{c}{\textbf{Pegasus~III}}\\ 
\multicolumn{1}{l}{Parameter} & \cite{Kim2015}          & \cite{Kim2016}                              & \multicolumn{1}{c}{This Work}               \\ \midrule
$M_V$ & $-4.1 \pm 0.5$ & $-3.4 \pm 0.4$ & $-4.17$ $^{+0.19}_{-0.22}$
\\\midrule
\multicolumn{4}{c}{\textit{Exponential}}\\
RA (h:m:s) & 22:24:22.6 $\pm 15\arcsec$ & 22:24:24.48 & 22:24:25.82 $\pm 5\arcsec$  \\
DEC (d:m:s) & +05:25:12 $\pm 14\arcsec$ & +05:24:18.0 & +05:24:54.01 $\pm 3\arcsec$ \\
$\theta_{\mathrm{exp}}$ (deg)          & $133 \pm$ 17            & $114^{+19}_{-17}$      & $85 \pm 8$   \\
$\epsilon_{\mathrm{exp}}$              & $0.46^{+0.18}_{-0.27}$ & $0.38^{+0.22}_{-0.38}$ & $0.36^{+0.09}_{-0.10}$ \\
$a_{h,\mathrm{exp}}$ (arcmin)          & $1.3^{+0.5}_{-0.4}$    & $0.85 \pm$ 0.22         & $1.88^{+0.42}_{-0.33}$ \\
$a_{h,\mathrm{exp}}$ (pc)          & $78^{+30}_{-24}$    & $53 \pm$ 14         & $118^{+31}_{-30}$ \\
\midrule
\multicolumn{4}{c}{\textit{Plummer}}\\
RA (h:m:s) & \multicolumn{1}{c}{-} & \multicolumn{1}{c}{-} & 22:24:25.78 $\pm 5\arcsec$\\
DEC (d:m:s) &  \multicolumn{1}{c}{-} & \multicolumn{1}{c}{-} & +05:24:54.17 $\pm 3\arcsec$\\
$\theta_{p}$ (deg)            & \multicolumn{1}{c}{-}   & \multicolumn{1}{c}{-}         & $83^{+8}_{-7}$                               \\
$\epsilon_{p}$                & \multicolumn{1}{c}{-}   & \multicolumn{1}{c}{-}         & $0.37^{+0.08}_{-0.09}$                            \\
$a_{h,p}$ (arcmin)            & \multicolumn{1}{c}{-}   & \multicolumn{1}{c}{-}         & $1.67^{+0.26}_{-0.21}$                     \\ 
$a_{h,\mathrm{exp}}$ (pc)              & \multicolumn{1}{c}{-}   & \multicolumn{1}{c}{-}         & $104^{+20}_{-23}$ \\
\hline
\end{tabular}
\caption{Absolute magnitude and structural properties for  Pegasus~III, with the top six lines reporting the absolute $V$-band magnitude and best-fit exponential values and the bottom five listing the best-fit Plummer values. The uncertainties reported for RA, DEC, $\theta$, $\epsilon$, and $a_h$ in arcminutes are the 16th and 84th percentiles from the MCMCs. The uncertainties on $M_V$ and $a_h$ in parsecs are the 16th and 84th percentiles from Monte Carlo simulations that took into account the errors on the distance modulus. \cite{Kim2015} and \cite{Kim2016} did not fit Plummer profiles to their data, and \cite{Kim2016} did not provide uncertainties on their central positions.}
\label{tab:peg3}
\end{table*}

\begin{table*}
\centering
\begin{tabular}{@{}lllll@{}}
\multicolumn{5}{c}{\textbf{Pisces~II}}\\ 
\multicolumn{1}{l}{Parameter} & \multicolumn{1}{c}{\cite{Belokurov2010}} & \multicolumn{1}{c}{\cite{Sand2012}} & \multicolumn{1}{c}{\cite{Munoz2018}} & \multicolumn{1}{c}{This Work}             \\ \midrule
$M_V$ & $-5.0 \pm 0.5$ & $-4.1 \pm 0.4$ & $-4.22 \pm 0.38$ & $-4.28^{+0.19}_{-0.16}$
\\\midrule
\multicolumn{5}{c}{\textit{Exponential}}\\
RA (h:m:s) & \multicolumn{1}{c}{-} & 22:58:32.33 $\pm 5\arcsec$ &  22:58:32.28 $\pm  9\farcs15$ & 22:58:32.76 $\pm 2\arcsec$ \\
DEC (d:m:s) & \multicolumn{1}{c}{-} & +05:57:17.7 $\pm 4\arcsec$  & +05:57:09.36 $\pm 5\farcs7$ & +05:57:20.36 $\pm 1\arcsec$\\
$\theta_{\mathrm{exp}}$ (deg)    &  \multicolumn{1}{c}{-}    & 107      & $98 \pm 13$                & $97 \pm3$                         \\
$\epsilon_{\mathrm{exp}}$        &  \multicolumn{1}{c}{-}    & \textless 0.28          & $0.39 \pm 0.10$     & $0.37 \pm 0.04$                    \\
$a_{h,\mathrm{exp}}$ (arcmin)    &  \multicolumn{1}{c}{-}    & $1.09 \pm 0.19$  & $1.18 \pm 0.20$            & $1.31^{+0.10}_{-0.09}$ \\ 
$a_{h,\mathrm{exp}}$ (pc)    &  \multicolumn{1}{c}{-}    & $58 \pm 10$  & $62.5 \pm 10.6$           & $69 \pm 8$  \\ 
\midrule
\multicolumn{5}{c}{\textit{Plummer}}\\
RA (h:m:s) & 22:58:31 $\pm 6\arcsec$ & 22:58:32.20 $\pm 5\arcsec$ & 22:58:32.28 $\pm  9\farcs15$ & 22:58:32.75 $\pm 2\arcsec$\\
DEC (d:m:s) & +05:57:09 $\pm 4\arcsec$ & +05:57:16.3 $\pm 4\arcsec$ & +05:57:09.36 $\pm 5\farcs7$ & +05:57:19.96 $\pm 1\arcsec$\\
$\theta_{p}$ (deg)      &     $77 \pm 12$      & $110 \pm 11$     & $78 \pm 20$         & $98 \pm 3$                         \\
$\epsilon_{p}$          &     $0.4 \pm 0.1$      & $0.33 \pm 0.13$  & $0.34 \pm 0.10$   & $0.37^{+0.03}_{-0.04}$                     \\
$a_{h,p}$ (arcmin)      &     $1.1 \pm 0.1$      & $1.12 \pm 0.18$  & $1.12 \pm 0.16$   & $1.34^{+0.08}_{-0.07}$                     \\ 
$a_{h,p}$ (pc)      &     $58 \pm 5$*      & $60 \pm 10$  & $59.3 \pm 8.5$   & $71 \pm 8$                     \\ 
\bottomrule
\end{tabular}
\caption{Same as Table \ref{tab:peg3}, for Pisces~II. \cite{Belokurov2010} did not fit an exponential model to their data. \cite{Sand2012} had unconstrained uncertainties for the position angle in their exponential model and their reported ellipticity is the 68\% upper confidence limit. *The uncertainty on the $a_h$ in parsecs was derived for this work from a Monte Carlo simulation using the uncertainty on the $a_h$ in arcminutes and the distance modulus of 21.3 reported in \cite{Belokurov2010}.}
\label{tab:psc2}
\end{table*}

The parameters of elliptical half-light radius and background surface density appear both correlated with the richness (number of stars) and with each other. The elliptical half-light radius has a positive correlation with the number of stars, while the background density has a negative correlation with the richness parameter.  
The elliptical half-light radius and background density parameters also have a negative correlation. 
One would expect the richness to increase with the elliptical half-light radius for a galaxy with the same central surface density ($S_0$), as $S_0$ is proportional to the richness divided by the squared scale radius, which in this case is the elliptical half-light radius. As the radius increases, for the $S_0$ to remain the same, the richness must also increase.
If the elliptical half-light radius is smaller, however, then a greater percentage of stars within the FOV would be expected to belong to the background. Similarly, for a given number of stars within a FOV, if a higher percentage of stars belongs to the galaxy, then a lower number will be attributed to the background.

\begin{figure*}
    \centering
    \includegraphics[width=\textwidth]{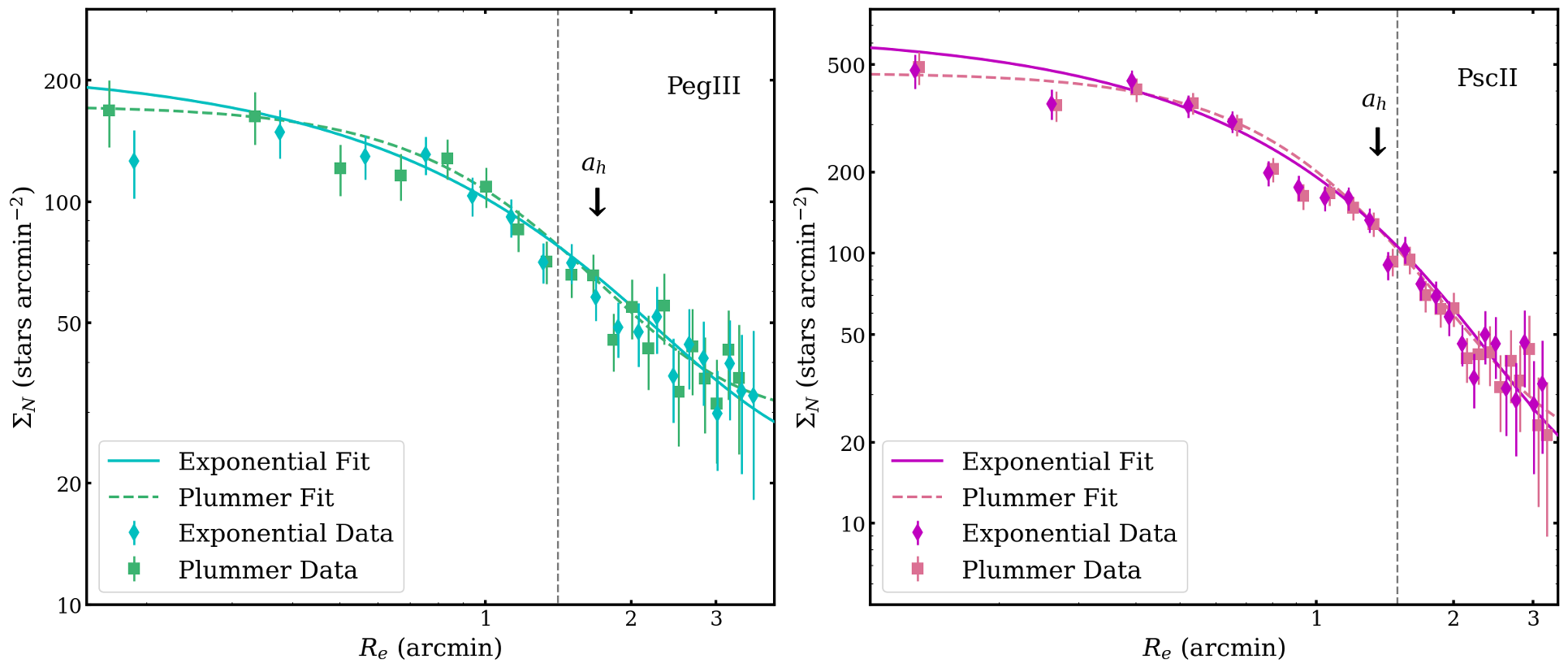}
    \caption{Best-fitting models for the surface density profiles of  Pegasus~III (left) and Pisces~II (right) plotted against the elliptical radius. 
    The curves show the best-fit one dimensional exponential (solid) and Plummer (dashed) profiles, which are fit to the full stellar distribution, not the annularly binned data shown here. The gray dashed line marks the approximate point where the elliptical annuli used for the surface density measurements begin to cover area outside the FOV in pixel space without source information in our catalogs. This was corrected for by dividing the number of stars in each elliptical annulus by only the area overlapping the FOV. The diamonds (squares) represent the surface density measurements in bins using the exponential (Plummer) model, with elliptical radii in increments of 0.1$a_h$. The errors come from a Poisson distribution. The $a_h$ marks the data points corresponding to the best-fit elliptical half-light radius.}
    \label{fig:profiles}
\end{figure*}

We show the best-fitting exponential and Plummer profiles for each galaxy (left: Peg~III, right: Psc~II) in Figure \ref{fig:profiles}. Surface density measurements taken at 0.1 increments of the elliptical half-light radius are plotted against the elliptical radius $R_e$ of each annular bin.

\begin{figure*}
    \centering
    \includegraphics[width=\textwidth]{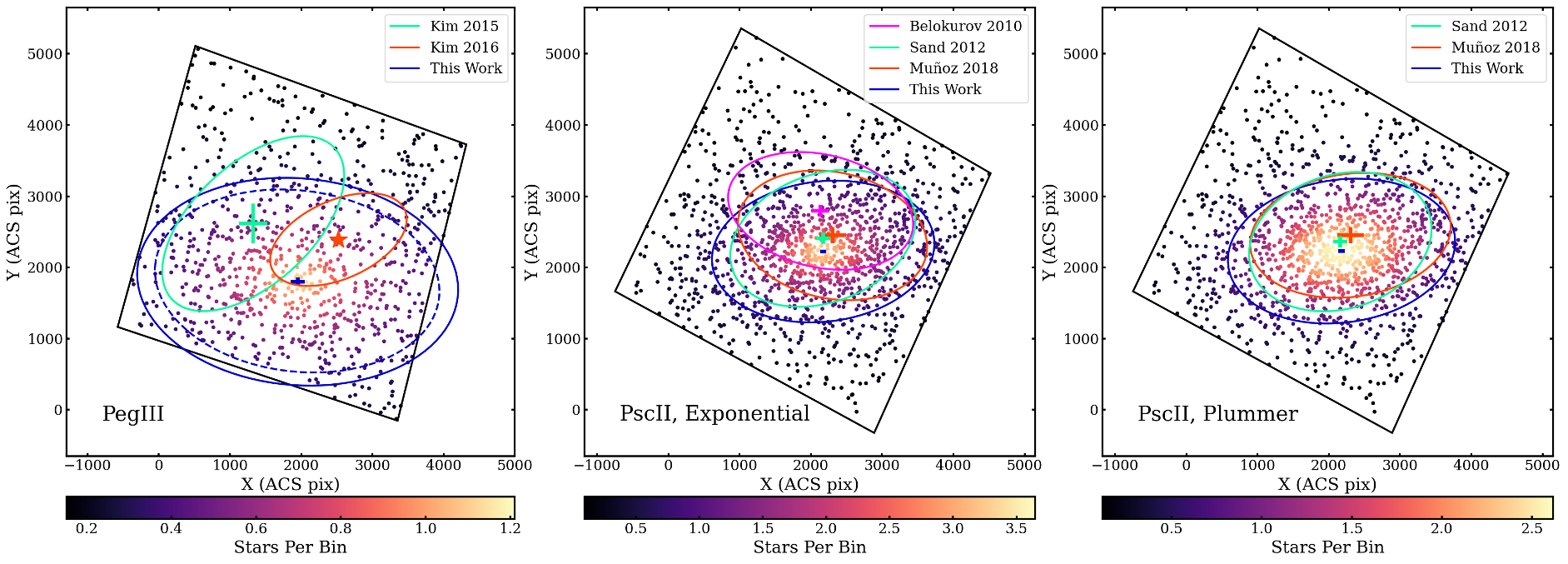}
    \caption{Comparison of past literature fits. Individual sources are colored based on the binned surface densities of the best-fitting exponential models in the left and center panels and the best-fitting Plummer model in the right panel. \textbf{Left}:  Pegasus~III. Exponential profile fits from \cite{Kim2015} and \cite{Kim2016} are shown in green and orange, respectively. The exponential profile fit from this work is shown in solid blue. The Plummer fit from this work is shown as the dashed blue ellipse. The central position is shown with error bars for \cite{Kim2015} in green and for this work's exponential fit in blue. The central position from this work's Plummer fit is not shown as it almost completely overlaps with the exponential value. The central position from \cite{Kim2016} is marked with an orange star, as no errors were reported. 
    \textbf{Middle}: Exponential profile fits and central positions with error bars for Pisces~II. \cite{Belokurov2010} is shown in fuchsia, \cite{Sand2012} in green, \cite{Munoz2018} in orange, and this work in blue.
    \textbf{Right}: Plummer profile fits, with the same assigned colors as in the middle panel.}
    \label{fig:combEllipse}
\end{figure*}

While the structural parameters we measured for Psc~II are consistent with previous literature values, the $a_h$ value of Peg~III (1\farcm88$^{+0.42}_{-0.33}$) is much larger ($\sim$2.2$\times$) than the most recent literature value from \cite{Kim2016} ($0\farcm85 \pm 0.22$, referred to as $r_h$ in their paper). This can be seen in the leftmost panel of Figure \ref{fig:combEllipse}. Here, we illustrate the best-fit 2D-exponential models projected onto the sources in our FOV. The best-fit Plummer model from this work is shown as the dashed ellipse. 
Both the \cite{Kim2015,Kim2016} structural fits show the semimajor axis extending in the SE to NW direction, while our fits extend more along the east to west axis. The NW overdensity in our Figure \ref{fig:contours} and smoothed data and residuals in Figure \ref{fig:dmrPeg3} could be have affected the \citeauthor{Kim2016} fits.
The same best-fit 2D-model comparisons are shown for Psc~II, with the exponential fits shown in the middle panel and the Plummer fits in the right panel.


As discussed in \cite{Munoz2012}, the FOV should be at least $3\times$ the half-light radius (assuming circular symmetry) in order to measure the morphology to 10\% accuracy. With the ACS FOV width of $202\arcsec$, or $3\farcm36$, we cover approximately 3.96$\times$ the $0\farcm85$ value for the elliptical half-light radius that we expected based on \cite{Kim2016}. 
However, when compared to our measured values for Peg~III, the ACS FOV is $\sim$1.8$\times$ our exponential $a_h$ and $\sim$2$\times$ our Plummer $a_h$. For Psc~II, the ratios are $\sim$2.6$\times$ and $\sim$2.5$\times$ for the exponential and Plummer $a_h$ values, respectively.
We explore trends in the structural parameters as a function of $a_h$ to FOV ratio in the next subsection.

\subsection{Mock Galaxy Tests}
\begin{figure}
    \centering
    \plotone{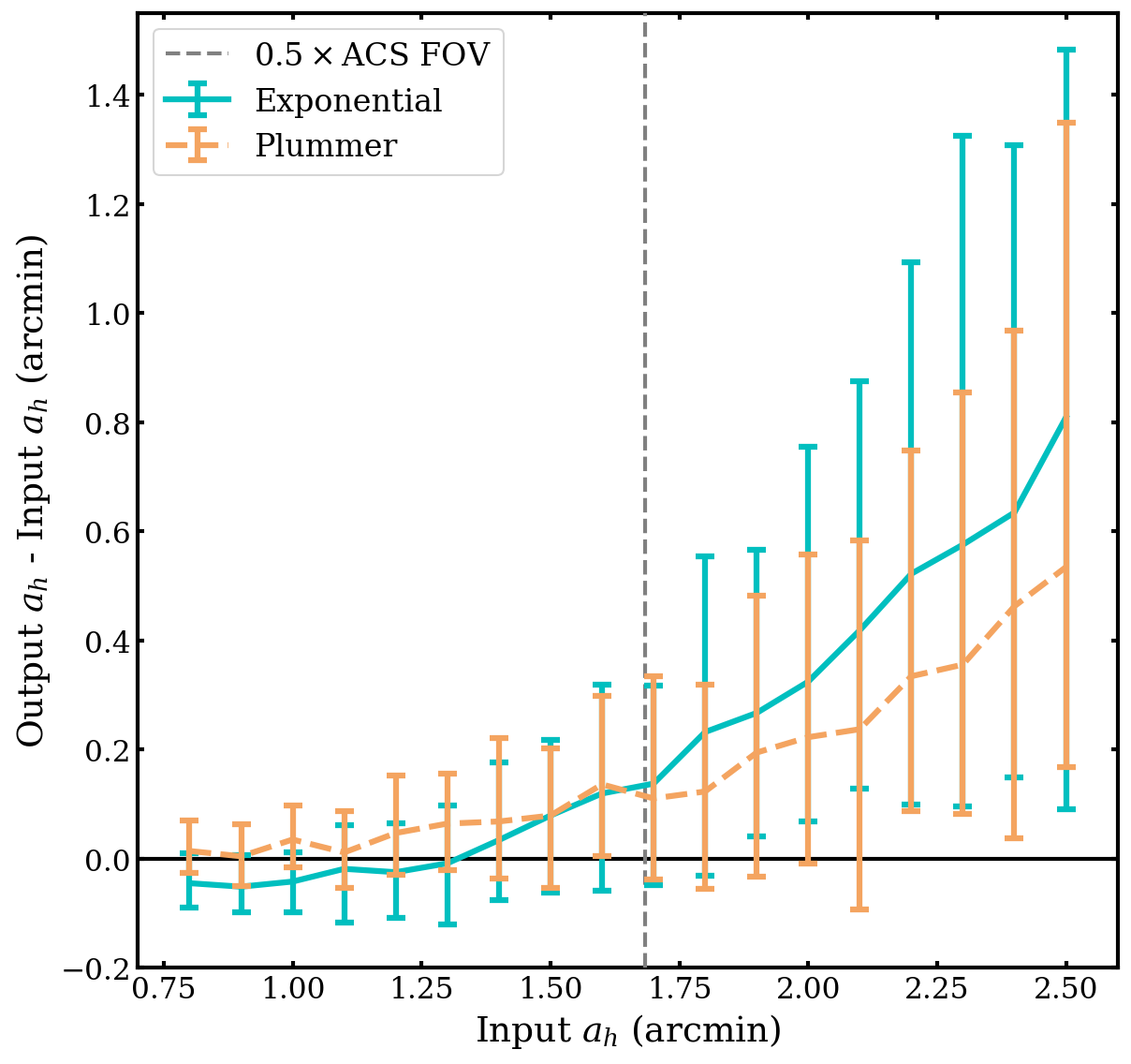}
    \caption{Average difference of the MCMC-fit output $a_h$ and set input $a_h$ for 100 mock galaxies. The exponential fit differences are shown in cyan and the Plummer in orange. The black horizontal line at 0 represents the ideal case, where the fit $a_h$ matches the input $a_h$. The error bars represent the 16th and 84th percentiles from the 100 fits. The analysis for each mock galaxy is performed over a FOV equivalent to that of the \textit{HST} ACS FOV, with the gray dashed vertical line marking where the input $a_h$ begins to be greater than one-half of the ACS FOV.}
    \label{fig:mockRh}
\end{figure}

To test how well our fitting procedure could recover the elliptical half-light radius for a galaxy given a limited FOV, we created simulated galaxies in Cartesian space based on the best-fit values for Peg~III. After generating distributions based on the probability density functions of the Plummer and exponential models, we randomly drew stars to match the length of the final Peg~III catalog and added a set number of background stars matching the best-fit background density value. These simulated data were then masked with the ACS FOV and chip gap and run through the MCMC. The results of these tests are shown in Figure \ref{fig:mockRh}.

From this, we see that the code can recover $a_h$ up to $\sim$1$\farcm4$ before it begins to return larger than the input $a_h$ values. Even up to $1\farcm8$ however, the expected $a_h$ values do fall within the uncertainties for both models, and remain close up through $2\farcm5$. Our measured $a_h$ values for Peg~III of $1\farcm88$ and $1\farcm67$ for the exponential and Plummer profiles, respectively, could therefore be taken as upper limits on the true value. The maximum well-recovered value of $\sim$1$\farcm4$ extending beyond the previously reported $a_h=0\farcm85$ \citep{Kim2016} suggests that our modeling would have been able to capture if Peg~III was indeed so compact.

\begin{table}
\centering
\begin{tabular}{@{}lll@{}}
\toprule
Parameter   & Peg~III & Psc~II \\ \midrule
$(m-M)_0$ & $21.66\pm0.12$ & $21.31\pm0.17$\\
$v_{\mathrm{GSR}}$ (km s$^{-1}$) & $-67.6 \pm 2.6$ & $-79.9 \pm 2.7$\\
$\sigma_v$ (km s$^{-1}$) & $5.4^{+3.0}_{-2.5}$ & $5.4^{+3.6}_{-2.4}$\\
$\mu_{\alpha}$cos$\delta$ (mas yr$^{-1}$) & $0.06 \pm 0.1$ & $0.11 \pm 0.11$ \\
$\mu_{\delta}$ (mas yr$^{-1}$) & $-0.2 \pm 0.1$ & $-0.24^{+0.12}_{-0.11}$\\ \midrule
$r_h$ (arcmin) &    $1.51^{+0.35}_{-0.29}$ &   $1.04 \pm 0.08$     \\
$r_h$ (pc)     &     $94^{+25}_{-24}$ &     $55 \pm 6$ \\
$m_V$          &    $17.50^{+0.15}_{-0.21}$ & $17.03 \pm 0.04$       \\
$M_{1/2}$ ($10^6$ M$_{\odot}$) & $3.2^{+4.3}_{-2.1}$  &  $1.9^{+3.3}_{-1.3}$  \\
($M/L_V$)$_{1/2}$ (M$_{\odot}/$L$_{\odot}$) &  $1600^{+480}_{-580}$  &  $850^{+570}_{-260}$\\
\hline
\end{tabular}
\caption{Adopted and derived values for Peg~III and Psc~II. The Peg~III distance modulus is from \cite{Kim2016}, and the distance modulus for Psc~II is from \cite{Sand2012}. The $\sigma_v$ values are from \cite{Kim2016} and \cite{Kirby2015} for Peg~III and Psc~II, respectively, while the PMs are both taken from \cite{mcconnachie21b}.
The azimuthally-averaged half-light radius $r_h$ is calculated using the relation $r_h{=}a_h \sqrt{1-\epsilon}$. ($M/L_V$)$_{1/2}$ is the mass-to-light ratio within the elliptical half-light radius. The processes for the derivation of $m_V$, $M_{1/2}$, and ($M/L_V$)$_{1/2}$ are described in Sections \ref{sec:mag} and \ref{sec:ML}.
The uncertainties on all derived quantities are the 16th and 84th percentiles of Monte Carlo simulations including the full-error space of all the relevant terms.}
\label{tab:derived}
\end{table}

\subsection{Apparent and Absolute V-Band Magnitude Calculations} \label{sec:mag}
To derive the integrated magnitudes of each UFD, we opted for a probabilistic model approach as opposed to using discrete stars.
To begin, we selected a box in color-magnitude space from $0\leq(V-I)\leq1.2$ and $20\leq m_V\leq29$, as this encompassed the majority of stars from the CMD that would be expected to belong to the galaxies.
For both the on-field and off-field, we created a Gaussian kernel using \texttt{scikit-learn} \citep{scikit-learn} \texttt{Kernel Density} and fit the kernel to the stars inside the color-magnitude box (CMD box). 
We gridded this space into 100 bins along each dimension, resulting in color bins $\sim$0.01 dex and magnitude bins $\sim$0.1 dex in width. We then generated random samples from the on- and off-field kernels and computed the log-likelihood of each sample under the model. The off-field was used to create a probabilistic background model from which we estimated the excess flux. Integrating in both color and magnitude space, we calculated the stellar density in the CMD box by multiplying the log-likelihoods by the area of the CMD box and subtracting the off-field model from the on-field. The integral returned the flux of the stars inside the CMD box. We multiplied this flux by a correction factor derived from the exponential models to account for the flux outside the FOV. Converting this flux back to magnitude space yielded the integrated apparent magnitude.

This calculation was performed within a Monte Carlo simulation (MC) that included the individual source magnitude errors and FOV corrections calculated from different sets of model parameters. We report the median integrated $m_V$ values of $17.50^{+0.15}_{-0.21}$ and $17.03\pm0.04$ for Peg~III and Psc~II, respectively, with the uncertainties representing the 16th and 84th percentiles. 
For the integrated $M_V$, we performed the same MC, this time also including errors on the distance modulus. We used the \cite{Kim2016} value of $21.66\pm0.12$ for Peg~III and the \cite{Sand2012} $21.31\pm0.17$ value for Psc~II. The most recent distance estimates \citep{Garofalo2021} indicate that the two UFDs have more similar heliocentric distances (Peg~III$\sim$174 kpc; Psc~II$\sim$175 kpc), however, our CMDs in Figure \ref{fig:cmd} suggest that Peg~III is farther away than Psc~II, as the horizontal branch and main-sequence turnoff for Peg~III are both $>$0.5 magnitudes less bright than those of Psc~II. As such, we chose to move forward with the previously reported literature values in our MC, which gave the median $M_V{=}-4.17^{+0.19}_{-0.22}$ for Peg~III and $-4.28^{+0.19}_{-0.16}$ for Psc~II, with the 16th and 84th percentiles quoted as the uncertainties.

\subsection{Mass-to-Light Ratios}
\label{sec:ML}
Using the newly derived elliptical half-light radii values, we calculate updated mass values using velocity dispersion values from \cite{Kirby2015} and \cite{Kim2016}. 
We use Equation \ref{eq:mass} derived by \cite{Wolf2010}, which was also employed by \cite{Kim2016} to estimate the mass within the elliptical half-light radius of Peg~III. $R_e$ is defined as the 2D-projected half-light radius from elliptical fits of surface brightness profiles and aligns with our $a_h$ values.

\begin{equation}
    M_{1/2} \simeq \frac{4}{G} \sigma^2_v R_e 
    \label{eq:mass}
\end{equation}

The \cite{Kim2016} $\sigma_v{=}5.4_{-2.5}^{+3.0}$ km s$^{-1}$, determined using seven member stars, and our elliptical half-light radius measurement of 118 parsecs at 215 kpc gives a $M_{1/2}{=}3.2^{+4.3}_{-2.1} \times 10^6$~M$_{\odot}$ for Peg~III. Converting our $M_V$ value to luminosity, we obtain the mass-to-light ratio within the elliptical half-light radius $(M/L_V)_{1/2}{=}1600^{+480}_{-580}$ M$_{\odot}/$L$_{\odot}$. This is within one sigma of the previously derived $1470^{+5660}_{-1240}$ M$_{\odot}/$L$_{\odot}$ from \cite{Kim2016}.
Using the \cite{Kirby2015} $\sigma_v{=}5.4^{+3.6}_{-2.4}$ km s$^{-1}$, which was also calculated using seven member stars, and our elliptical half-light radius of 69 parsecs at 183 kpc, we calculate a $M_{1/2}{=}1.9^{+3.3}_{-1.3} \times 10^6$~M$_{\odot}$ for Psc~II. The $(M/L_V)_{1/2}$ is then $850^{+570}_{-260}$~M$_{\odot}/$L$_{\odot}$. This agrees within two sigma with the value of $370^{+310}_{-240}$~M$_{\odot}/$L$_{\odot}$ that \cite{Kirby2015} derived.

\section{Orbital Analysis}
\label{sec:orbits}

\subsection{Previous Literature}
\cite{Kim2015} and \cite{Kim2016} hypothesized a connection between Peg~III and Psc~II immediately upon the discovery of the former, due at first to their small on-sky separation (8.5$^{\circ}$), line-of-sight separation of $\sim$32~kpc, and similar heliocentric distances, and later from their similar radial velocities and calculated physical separation of $\sim$36~kpc.
\cite{Kim2016} also noted the presence of an irregularity in the isodensity lines for Peg~III in the direction of Psc~II, but included the caveat of small number statistics.

Focusing on the ellipticity of Peg~III, which we find to be $0.36^{+0.09}_{-0.10}$, \cite{Kim2016} considered that, if not caused by the formation process, this could be the result of tidal interactions with the MW, a more likely scenario if Peg~III is not in dynamic equilibrium or if unresolved binaries are inflating the velocity dispersion. In such a case, they suggested that Peg~III might be a tidally disrupted dwarf galaxy remnant but would need to have had a highly eccentric orbit to achieve close enough Galactocentric distances to experience tidal effects. Alternatively, the ellipticity of Peg~III could have been caused by a tidal interaction with Psc~II, though at the time, there was not enough orbital information for \cite{Kim2016} to further investigate this possibility. 
While our mass estimate for Peg~III is greater than that of Psc~II, \cite{Kim2016} measured Peg~III to be $0\farcm85$ and thus smaller than Psc~II, which could explain why they suggested Peg~III could have been tidally affected by Psc~II.

\subsection{Proper Motion Measurements}
Recently, several authors have measured the proper motions (PMs) of Peg~III and Psc~II using data from \textit{Gaia} eDR3 \citep{mcconnachie21b,Li21, Battaglia21}, now making it possible to explicitly calculate the orbital histories of both galaxies simultaneously to assess whether or not they could be or have been a bound pair. 

\cite{Li21} reported a PM for Psc~II using three stars and found a highly eccentric orbital history such that Psc~II only reaches its closest distance relative to the MW at its current distance. By integrating both forward and back for 10 Gyr, they also concluded that Psc~II shows a high probability of being unbound from the MW's halo (95.85\%). 

\cite{Battaglia21} independently measured the PM of Psc~II using two member stars from \textit{Gaia} eDR3, and though they measured a different PM for Psc~II and included the infall of a massive LMC in their models, their measurements still implied that Psc~II is not and has never been bound to the MW. Both \citet{Battaglia21} and \citet{Li21} tested a similar range of MW mass models to arrive at these conclusions.

\cite{mcconnachie21b} reported the first PM value for Peg~III measured using only one member star, in addition to their own estimate for the PM of Psc~II using four spectroscopic member stars, three of which had radial velocities more than 3$\sigma$ away from the mean velocity of the system. The reported uncertainties were smaller than those of \citet{Li21} and \citet{Battaglia21} by a factor of two or more. However, unlike \citet{Li21} and \citet{Battaglia21}, \citet{mcconnachie21b} imposed a prior on their proper motion determinations such that the galaxies are assumed to be bound to the MW (in direct contrast to the orbital histories suggested by the \citet{Battaglia21} and \citet{Li21} PMs).
Given that Peg~III and Psc~II are outer MW dwarfs, such a restrictive prior biases the measured PMs towards zero. In this work, we calculated orbital histories using the \citet{mcconnachie21b} PMs (see Table \ref{tab:derived}) for both galaxies given the uniform methodology that yields PMs with the lowest reported uncertainties for both dwarfs, keeping in mind that the prior could affect the resulting orbital histories and subsequent interpretations.\footnote{Following the initial submission of this paper, \cite{pace2022} measured new systemic proper motions for Peg~III and Psc~II using the structural parameters reported herein. Their measurements agree with the \cite{mcconnachie21b} values used in our orbital analysis within one to two sigma.}  

\subsection{Orbital Model and Results}
\begin{table*}
\begin{center}
\begin{tabular}{lcccccc}\\
\multicolumn{7}{c}{\textbf{MW only potential}}\\ \hline
{Orbit} & {$\rm f_{peri}$ } & {$\rm t_{peri}$ [Gyr]} &{$\rm r_{peri}$ [kpc]} &  {$\rm f_{apo}$} & {$\rm t_{apo}$ [Gyr]} & {$\rm r_{apo}$ [kpc]}   \\ \hline \hline
\multicolumn{7}{c}{MW1} \\ \hline
PegIII-MW & 0.53 & 2.56 [2.61,4.53]  &  20 [24,115] & 0.67 &  0.58 [0.64,2.61] &   217 [205,279] \\ 
PscII-MW &  0.58 & 2.56 [2.54,4.65]  &  9 [22,103] & 0.70 &  0.71 [0.71,2.42] &   203 [196,270] \\ 
PegIII-PscII & 0.75 & 1.8 [0.06,3.65]  &  12 [20,113] & 0.61 &  2.4 [2.21,4.63] &   17 [159,572] \\\hline

\multicolumn{7}{c}{MW2} \\ \hline
PegIII-MW & 0.7 & 2.07 [2.14,4.12]  &  15 [26,130] & 0.82 &  0.42 [0.46,2.24] &   213 [201,278] \\ 
PscII-MW &  0.73 & 2.06 [2.05,3.85]  &  11 [22,112] & 0.84 &  0.5 [0.51,2.23] &   197 [189,269] \\ 
PegIII-PscII & 0.84 & 2.1 [0.07,3.47]  &  7 [22,128] & 0.82 &  5.31 [1.8,4.05] &   256 [146,523] \\  \hline
\end{tabular}
\caption{Orbital parameters for Peg~III and Psc~II using the \citet{mcconnachie21b} proper motions for both galaxies.  The orbital parameters listed here are for the representative cases shown in Figure \ref{fig:orbits}, rather than the median orbital history. The listed uncertainties correspond to the 16th and 84th percentiles around the median of the distributions calculated using 1,000 Monte Carlo samples in the MW-only potential to illustrate the range of orbital uncertainty. $\rm f_{peri}$ ($\rm f_{apo}$) is the fraction of 1,000 orbits where a pericenter (apocenter) is recovered, $\rm t_{peri}$ ($\rm t_{apo}$) is the time at which the pericenter (apocenter) occurs on average, and $\rm r_{peri}$ ($\rm r_{apo}$) is the distance of the most recent pericenter (apocenter). Pericenter and apocenter are defined as the critical minima and maxima along orbital trajectories in Galactocentric distance.}
\label{tab:MWonly}
\end{center}
\end{table*}

\begin{table*}
\begin{center}
\begin{tabular}{lcccccc}\\
\multicolumn{7}{c}{\textbf{MW-LMC potential}}\\ \hline
{Orbit} & {$\rm f_{peri}$ } & {$\rm t_{peri}$ [Gyr]} &{$\rm r_{peri}$ [kpc]} &  {$\rm f_{apo}$} & {$\rm t_{apo}$ [Gyr]} & {$\rm r_{apo}$ [kpc]}   \\ \hline \hline
\multicolumn{7}{c}{MW1} \\ \hline
PegIII-MW & 0.71 & 1.88 [1.81,3.12]  &  95 [29,145] & 0.76 &  0.3 [0.3,0.74] &   211 [194,239] \\ 
PegIII-LMC & 0.71 & 0.85 [1.81,3.12]  &  32 [29,145] & 0.76 &  5.92 [0.3,0.74] &   817 [194,239] \\ 
PscII-MW &  0.72 & 2.08 [1.75,3.3]  &  99 [31,125] & 0.81 &  0.35 [0.33,1.05] &   193 [182,225] \\ 
PscII-LMC & 0.72 & 0.81 [1.75,3.3]  &  31 [31,125] & 0.81 &  4.8 [0.33,1.05] &   685 [182,225] \\ 
PegIII-PscII & 0.8 & 1.16 [0.06,2.7]  &  10 [21,97] & 0.75 &  2.18 [1.4,4.05] &   42 [106,441]\\ \hline

\hline
\multicolumn{7}{c}{MW2} \\ \hline
PegIII-MW & 0.8 & 1.64 [1.59,2.64]  &  55 [31,150] & 0.89 &  0.25 [0.25,0.82] &   210 [193,248] \\ 
PegIII-LMC & 0.8 & 0.85 [1.59,2.64]  &  56 [31,150] & 0.89 &  3.35 [0.25,0.82] &   476 [193,248] \\ 
PscII-MW &  0.82 & 1.64 [1.5,2.79]  &  66 [32,136] & 0.91 &  0.29 [0.28,1.22] &   191 [180,227] \\ 
PscII-LMC & 0.82 & 0.81 [1.5,2.79]  &  49 [32,136] & 0.91 &  3.38 [0.28,1.22] &   464 [180,227] \\
PegIII-PscII & 0.88 & 1.16 [0.07,2.84]  &  18 [21,113] & 0.9 &  1.7 [1.28,3.49] &   21 [109,440] \\ 
\hline
\end{tabular}
\caption{Same as Table \ref{tab:MWonly} but for the MW-LMC potential. Here, the orbital parameters of Peg~III and Psc~II calculated relative to the LMC's orbit are also included.}
\label{tab:MWLMC}
\end{center}
\end{table*}
Here, we explore whether the two UFDs could have interacted with each other, with the MW, or with the LMC. Radial velocities are adopted from \citeauthor{Kim2016}~(\citeyear{Kim2016}; Peg~III) and \citeauthor{Kirby2015}~(\citeyear{Kirby2015}; Psc~II) and distance moduli\footnote{During the course of the analysis for this paper, new distance measurements were published by \cite{Garofalo2021}, derived using RR Lyrae. The range of errors we considered in the distance moduli for the orbital analysis spans a large portion of their values of $174 \pm 18$ kpc for Peg~III and $175 \pm 11$ kpc for Psc~II, 
thus we did not perform separate calculations here because the PM error space is by far the dominant source of measurement uncertainties.} are adopted from \citeauthor{Kim2016}~(\citeyear{Kim2016}; Peg~III) and 
\citeauthor{Sand2012}~(\citeyear{Sand2012}; Psc~II), respectively.
Direct orbital histories are calculated using initial phase space coordinates that equal the direct transformation of PMs, LOS velocity, and distance converted to Galactocentric phase-space coordinates. 

Following the methodology of \cite{Patel20}, we integrate orbits for Peg~III and Psc~II in a three-component (Hernquist bulge + Miyamoto-Nagai disk + NFW halo) MW-only potential and a combined three-component MW and two-component (Miyamoto-Nagai disk + NFW halo) LMC potential using a fixed LMC mass of $1.8\times10^{11}$~M$_{\odot}$, 
which was used as the fiducial model in \cite{Patel20} and is consistent with abundance-matched halo estimates and recent Magellanic system models \citep[e.g.,][]{Guo2011,Besla2012,Besla2013,Besla2016} based on empirical measurements and dynamical mass arguments \citep[e.g.,][]{Kim1998,Majewski2009,Saha2010,vdMarel2014,Mackey2016}.
Here, the MW is allowed to move in response to the LMC's gravitational influence. Two MW masses are considered:\ MW1 has a virial mass of $10^{12}$~M$_{\odot}$, while MW2 has a virial mass of $1.5 \times 10^{12}$~M$_{\odot}$. Peg~III and Psc~II are modeled as Plummer spheres with a total mass of $10^9$~M$_{\odot}$ and a Plummer scale radius of 1~kpc \citep{Jeon2017}. 
Dynamical friction owing to the MW and the LMC is also included, as in \citet{Patel20}. See \citet{Patel20} Tables 3 and 4 for all MW and LMC model parameters.

One important difference with respect to the methodology of \citet{Patel20} is that we have explicitly accounted for the gravitational forces that Peg~III and Psc~II exert on each other since it has been speculated that they may be a bound pair. Thus, we calculate the joint, 4-body orbital history of the MW, LMC (when applicable), Peg~III, and Psc~II. We have also checked whether including the additional gravitational influence of the SMC is significant and find that the SMC negligibly affects the orbits of Peg~III and Psc~II, thus we do not incorporate it in the rest of this analysis.

\begin{figure*}
    \centering
    \includegraphics[width=0.7\textwidth, trim=0mm 5mm 0mm 0mm]{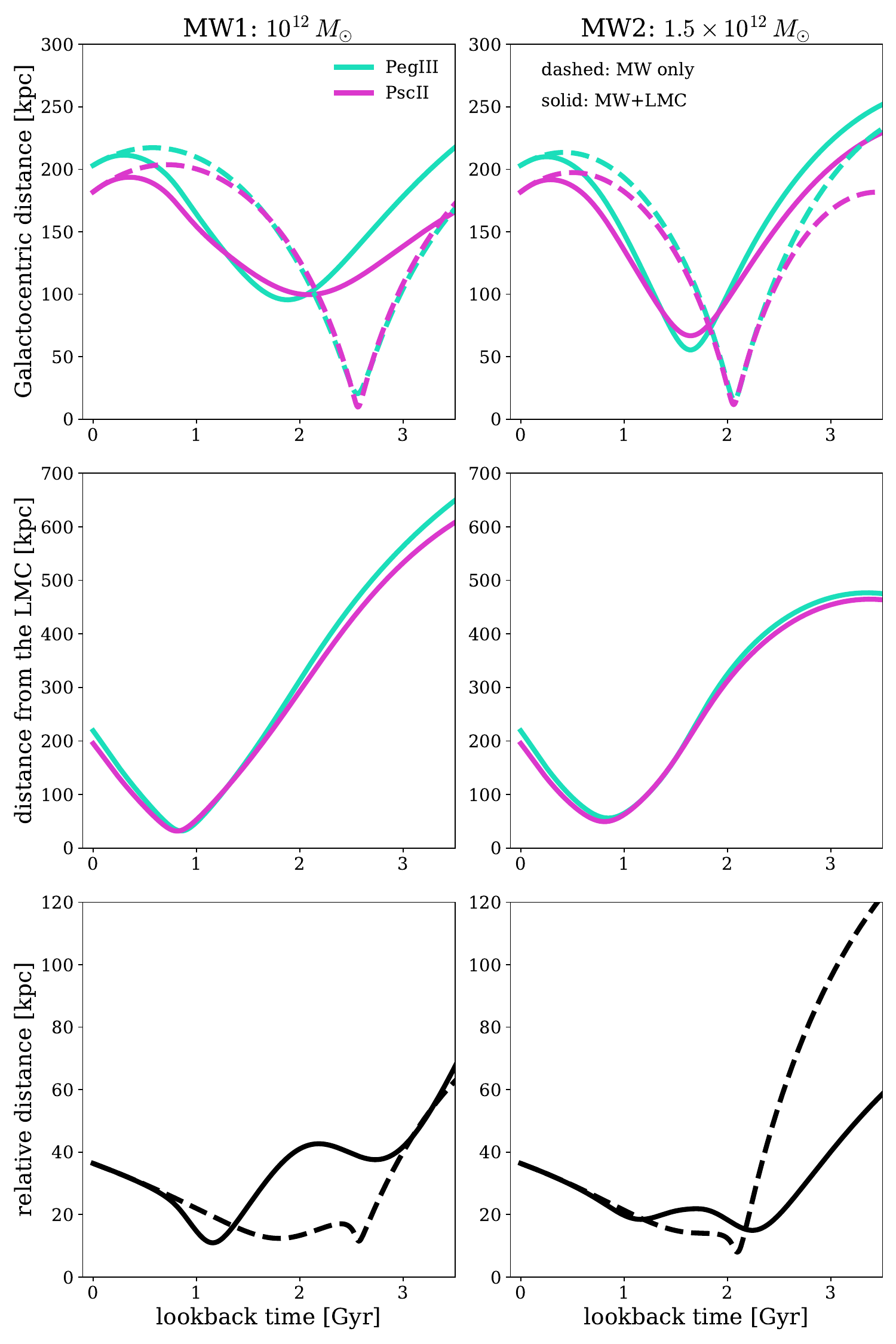}

    \caption{\textbf{Top:} Direct orbital histories of Peg~III (cyan) and Psc~II (magenta) relative to the MW, the LMC, and to each other are shown in the top, middle, and bottom panels, respectively. \textit{Gaia} eDR3 proper motions from \cite{mcconnachie21b} are used for both Peg~III and Psc~II. MW1 (left) has a virial mass of $10^{12}$~M$_{\odot}$ while MW2 (right) has a virial mass of $1.5 \times 10^{12}$~M$_{\odot}$. Dashed lines represent orbits computed in a MW-only potential and solid lines indicate a MW+LMC potential with an LMC mass of $1.8\times 10^{11}$~M$_{\odot}$. The presence of the LMC decreases the distance at pericenter relative to the MW-only model by $\sim$40--80~kpc. \textbf{Middle:} The orbital history of each UFD relative to the LMC. For both MW1 and MW2, the satellites complete a percentric passage about the LMC at $\sim$1~Gyr ago. \textbf{Bottom:} The orbit of Peg~III and Psc~II relative to one another. In both MW potentials, when the LMC is included, the two UFDs remain within $\sim$60~kpc of each other in the last 4~Gyr, indicating that the satellites may be an interacting pair that could have originated in a similar environment.}
    \label{fig:orbits}
\end{figure*}

The direct orbital histories of Peg~III (cyan) and Psc~II (magenta) using the \cite{mcconnachie21b} PM values are shown in Figure \ref{fig:orbits}. These orbits do not represent the uncertainties on the measured quantities (i.e., distance, proper motion, LOS velocity). Uncertainties on the orbital parameters that do account for such are provided in Tables \ref{tab:MWonly} and \ref{tab:MWLMC}. 

Orbital histories with respect to the MW (top panel, Figure \ref{fig:orbits}) in the MW-only potential (dashed lines) indicate that both Peg~III and Psc~II complete a very close ($\leq$20~kpc), pericentric approach around the MW at $\sim$2.5~Gyr (MW1) or $\sim$2~Gyr ago (MW2). However, when the influence of the LMC is included, the satellites complete a pericentric approach around the MW at $\sim$1.5~Gyr ago in both MW mass models at distances of $\sim$50--90~kpc. Thus, the influence of the LMC significantly increases the distance at pericenter about the MW and changes the timing at which the pericenter occurs. This effect is two-fold as it is caused by the gravitational influence of the LMC itself in addition to the LMC causing the MW's center of mass to move in response to the passage of the LMC. We also test LMC masses in the range 8--$25 \times 10^{10}$~M$_{\odot}$ and find no significant differences in the resulting orbital histories.

These results are in contrast to the role the LMC plays on the orbit of Crater~2, for example, where including the LMC plunges Crater~2 further into the halo of the MW, providing a possible explanation for its puffed up morphology \citep{Torrealba2016,Caldwell2017,Fritz2018,Fu2019,Ji2021}. 
Peg~III and Psc~II are much more compact than Crater~2, however, and are thus much less vulnerable to tidal effects. In upcoming work, we will further disentangle which physical effects associated with the passage of the LMC (i.e., the formation of the DM wake, the large-scale density perturbations introduced to the MW DM halo, or LMC tidal debris) are most significant for altering the orbital histories of low mass satellite companions in the MW's halo (Patel et al, in prep.). 
Indeed, in our density contour maps (Figure \ref{fig:contours}) and in the recent \cite{Garofalo2021} paper, there appear to be no signs of tidal effects, and while our Peg~III elliptical half-light radius is larger than previous literature values, it is still an order of magnitude smaller than that of Crater~2 ($\sim$118~pc compared to $\sim$1100~pc; \citealt{Torrealba2016}).

The middle panels of Figure \ref{fig:orbits} show the orbits of Peg~III and Psc~II with respect to the LMC as the LMC also falls into the halo of the MW. The orbits of the two UFDs are very similar to each other and exhibit a pericentric passage about the LMC at about 0.8~Gyr ago at $\sim$30~kpc for MW1 and $\sim$60~kpc for MW2. This yields further evidence that the dynamics of the UFDs have been strongly impacted by the gravitational influence of the LMC, even if they are not in close proximity to the LMC at present (Peg~III is 219~kpc from the LMC and Psc~II is 195~kpc from the LMC). It is unclear whether the satellites were ever bound to the LMC based on the calculations presented here. Note that for MW1, the LMC's median infall time is approximately 1.4 Gyr ago and about 7 Gyr ago for MW2 \citep{Patel17}. Despite this large variation in infall time for the LMC, the UFDs behave fairly similarly over the last $\sim$3 Gyr in both MW mass potentials. 

In addition to computing the direct orbital histories, we also assess how much the measurement uncertainties in PM, LOS velocity, and distance affect the corresponding orbital histories for Peg~III and Psc~II. To this end, we calculate 1,000 orbits for each orbital model: light MW-only, heavy MW-only, light MW+LMC, heavy MW+LMC. These 1,000 orbits use initial conditions drawn in a Monte Carlo fashion from the 1$\sigma$ uncertainty on PM, LOS velocity, and distance for each dwarf galaxy (plus the LMC when included), thus the phase space uncertainties for both galaxies are jointly sampled in the 3-body and 4-body calculations. 

In Tables \ref{tab:MWonly} and \ref{tab:MWLMC} we list the distance and timing of pericenter and apocenter for the representative direct orbital histories in Figure \ref{fig:orbits}. We use the direct values because the mean and median obtained from the Monte Carlo distribution can be biased towards larger values, especially for the apocenter \citep[see][]{Fritz2018}. The uncertainties provided correspond to the 16th and 84th percentiles around the median of the distributions resulting from 1,000 orbit calculations spanning the uncertainty in phase space coordinates. We report orbital parameters for each satellite with respect to the MW, LMC, and each other for both MW mass models. Note that orbital parameters are only listed for the fraction of orbits where a pericenter\footnote{Note that some studies will adopt the minimum relative distance when no critical minimum is recovered \citep[e.g.,][]{Battaglia21} however, we do not follow this methodology as the minimum distance does not always correspond to a true pericentric passage.} (or apocenter) is recovered along the orbital trajectory. Therefore, the uncertainties do not encompass the orbital histories where no such critical minima and maxima are recovered in the last 6~Gyr.

\subsection{Caveats}
As we do not account for the evolving MW potential or large scale structure, we only integrate orbits for the last 6~Gyr. Furthermore, it has been shown that the LMC significantly perturbs the shape and density of the MW's dark matter halo, loses a significant portion of its own mass, and creates a trailing wake of dark matter over the last $\sim$2~Gyr \citep{GC19,GC21}. Thus, the most physically accurate model of Peg~III and Psc~II's orbital histories would be calculated in a time-evolving MW+LMC potential accounting for such features \citep[][Patel, in prep.]{dsouza22}. Given the significant PM uncertainties and the complex dynamics of the MW-LMC system, we note that the orbits presented here are only an approximation of the most plausible orbital histories within the measured phase space of these dwarf galaxies, assuming they are indeed bound to the MW \citep[as imposed by the prior in][]{mcconnachie21b}.


\begin{figure*}
    \centering
    \includegraphics[width=\textwidth]{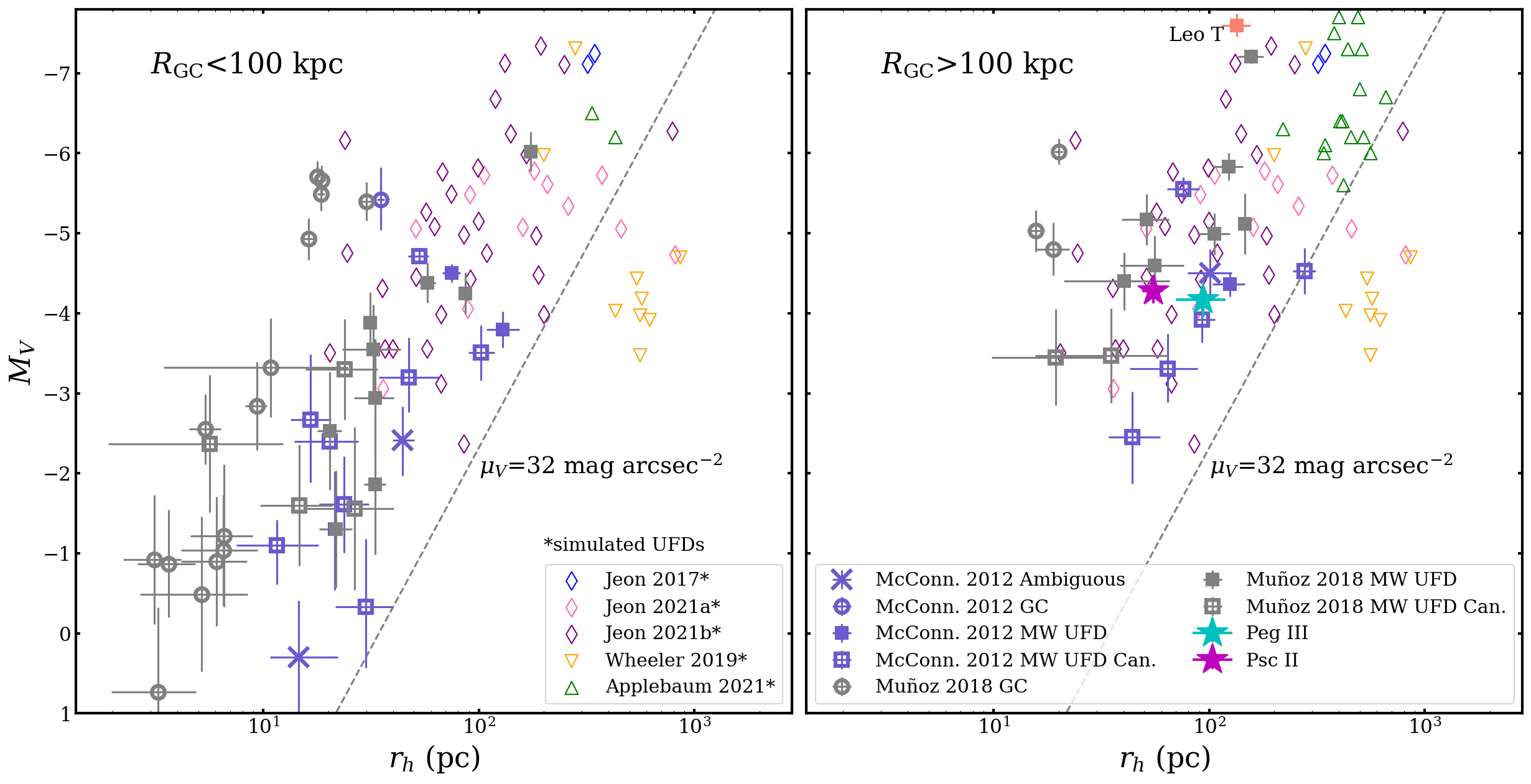}
    \caption{Comparison of simulated UFDs to observed MW satellites. The simulated UFDs (\citealt{Jeon2017}, blue diamonds; \citealt{Jeon2021a}, pink diamonds; \citealt{Jeon2021b}, purple diamonds; \citealt{Wheeler2019}, orange inverted triangles) are the same in both panels, except for the \cite{Applebaum2021} (green triangles), which are split between the two panels based on their simulated Galactocentric distances. Simulated UFDs can be distinguished from observed satellites as the latter all have error bars.
    The observed data are also split based on their Galactocentric distance, with satellites within 100 kpc shown in the left panel and satellites beyond 100 kpc in the right.
    Confirmed MW UFDs are shown as filled squares, candidate MW UFDs as open squares, MW globular clusters (GCs) as open circles, and ambiguous MW satellites as x's. The purple symbols use data from the updated \cite{McConnachie2012} table, while the gray symbols (and Leo T, in orange) are from \cite{Munoz2018}. The dashed line represents a constant surface brightness of 32 mag arsec$^{-2}$, approximately the current observational limit. Our measured Peg~III (cyan star) and Psc~II (magenta star) half-light radius and $M_V$ values lie in the right panel, within the $r_h$ and $M_V$ range of other MW UFDs as well as the \cite{Jeon2021a,Jeon2021b} simulated field UFDs. Some inner-halo satellites from the left panel also fall within the \cite{Jeon2021a,Jeon2021b} range.}
    \label{fig:rhmv}
\end{figure*}


\section{Discussion} \label{sec:disc}
\subsection{Comparison to Simulations and Other UFDs}
Peg~III and Psc~II are among the few known UFDs in the outer halo of the MW, at Galactocentric distances of $\sim$213 and $\sim$182 kpc, respectively. 
To see how our measured Peg~III and Psc~II azimuthally-averaged half-light radius ($r_h$) and $M_V$ values compare with those of other faint MW satellites, we place them in the size-luminosity plane (Figure \ref{fig:rhmv}). In addition to observed MW satellites (with $a_h$ converted to $r_h$ where necessary; \citealt{McConnachie2012,Munoz2018}), we have also included values from five sets of simulated UFDs: \cite{Jeon2017,Jeon2021a,Jeon2021b}, \cite{Wheeler2019}, and \cite{Applebaum2021}. The observed satellites are split between two panels to explore the difference between those closer (left; $<$100~kpc) and farther (right; $>$100~kpc) from the Galactic Center. 
The simulated galaxies are shown in both panels, except for the \cite{Applebaum2021} values, which are split between the two according to the Galactocentric distances determined within their framework.

To derive $M_V$ values for the \citeauthor{Jeon2017}\ and \cite{Wheeler2019} simulations, we used \texttt{Starburst99} \citep{Leitherer1999} to convert from stellar mass, based on the Padova evolutionary tracks \citep{Bressan1993,Fagotto1994a,Fagotto1994b,Girardi2000}, and the given half-stellar-mass radii as (circular) half-light radii. The half-light radii that \cite{Applebaum2021} report are circular and derived from the summation of particle luminosities.

The \cite{Wheeler2019} simulations were unable to produce UFDs with half-light radii lower than 200~pc, which the authors suggested could be in tension with current observations because telescopes might only be sensitive to the ``bright'' cores of diffuse and relatively massive objects. Similarly, the \cite{Applebaum2021} simulations did not produce any galaxies in the UFD magnitude range with smaller $r_h$ than $\sim$300~pc, which may have been due to their force softening.
Among the simulations included in our comparison, the \cite{Wheeler2019} and \cite{Jeon2021a,Jeon2021b} have the highest resolution, with \citeauthor{Wheeler2019} using $m_{\mathrm{gas}}{=}30$ and $m_{gas}{=}250$~M$_{\odot}$ in their high- and median-resolution simulations, respectively, and \citeauthor{Jeon2021b}\ using $m_{\mathrm{gas}}{\sim}60$~M$_{\odot}$.
As simulation techniques improve and are able to resolve a broader diversity of UFDs in a MW-environment, it will be interesting to see whether they more closely reproduce the scatter shown by \cite{McConnachie2012} and \cite{Munoz2018}. 

For Figure \ref{fig:rhmv}, we have imposed an upper magnitude limit of $M_V{=}{-}7.7$ to only include simulated galaxies in the UFD range, as \cite{Jeon2017}, \cite{Wheeler2019}, and \cite{Applebaum2021} produced galaxies in the dSph regime as well. As seen in Figure \ref{fig:rhmv}, Peg~III and Psc~II are well within the observed $M_V$ and $r_h$ ranges for other MW UFDs. They also fall in the range covered by the \cite{Jeon2021a,Jeon2021b} simulated field UFDs. Some closer MW UFDs and MW UFD candidates also fall within the simulated field UFD area, suggesting against this agreement being unique to outer-halo UFDs. Additionally, there is a higher number of observed satellites with smaller $r_h$ and fainter $M_v$ within 100~kpc, likely due to observational constraints. 

As new observatories come online with deeper detection limits, smaller and fainter satellites beyond 100~kpc could be discovered that have no analogs in current simulations. One might expect outer-halo satellites that have never been within 100~kpc of the MW to be more compact than UFDs with similar masses that have been closer and possibly subjected to strong tidal forces. If there are no or very few distant UFDs found in this smaller and fainter regime, the impact of the LMC and how it might have drawn UFDs closer ($\sim$50~kpc) at some point in their orbital history could be considered. While our orbital histories for Peg~III and Psc~II show the LMC increasing the pericenter distance, it is possible that other MW satellites (e.g., Crater~2, \citealt{Ji2021}) could have experienced very different effects.


\subsection{A Bound Pair?}
\label{subsec:bound}
Several proposed pairs of galaxies have been reported within the Local Group \citep{Pawlowski21}. 
Speculation that Peg~III and Psc~II are associated began in the Peg~III discovery paper \citep[][]{Kim2015}, due to their on-sky proximity and spatial separation of $\sim$30~kpc. The support for this association increased with the \cite{Kim2016} measurement of a radial velocity for Peg~III within $\sim$10~km~s$^{-1}$ of that of Psc~II. Additionally, they found the same irregularity in structure in their deeper imaging as in the discovery paper. 

The next work to investigate the possible connection between Peg~III and Psc~II was \cite{Garofalo2021}. Using LBT data, they found no irregular shape and concluded that there were no signs of past tidal interactions. In our density contour maps (Figure \ref{fig:contours}), we find no signs of an irregular shape for Peg~III 
(unlike \citealt{Kim2015,Kim2016}) or
for Psc~II, which we calculated to be the less massive of the two UFDs and therefore possibly more susceptible to tidal effects.
While previously limited to only looking at the morphologies for hints of past interactions, \textit{Gaia} eDR3 has provided us with the opportunity to use kinematics to study the UFDs' possible shared history. With proper motions in hand, Peg~III and Psc~II are amongst the first pairs of UFDs to be investigated further in this context.

To calculate a rough estimate of the likelihood that two satellite galaxies at similarly large distances and with radial velocities comparable to Peg~III and Psc~II would appear together by chance (i.e., not associated with each other prior to entering the MW halo) with a similar relative distance and radial velocity difference between them, we drew 1 million UFDs from a Gaussian velocity distribution centered around 0~km~s$^{-1}$ with $\sigma{=}100$~km~s$^{-1}$ and a distance distribution of $r^{3-1-\gamma}$ from 30 to 300~kpc, with $\gamma{=}2.11$ taken from \cite{Fritz2020}. We compared the drawn distance and velocity values to those of Psc~II, calculating the percentage of these UFDs with $\Delta v\leq15$~km~s$^{-1}$ and $\Delta r\leq40$~kpc with respect to the properties of Psc~II (${\sim}{-}80$~km~s$^{-1}$; $\sim$183~kpc). Of the 1 million draws, 2.5\% met the criteria, suggesting a low probability that two galaxies with such similar properties to each other as Peg~III and Psc~II share would exist in the MW halo by chance.

Furthermore, we can use the orbital histories computed using the \citet{mcconnachie21b} PMs to speculate about whether Peg~III and Psc~II have a shared orbital history or one that indicates that they have been bound in the recent past. The bottom panel of Figure \ref{fig:orbits} shows the orbits of Peg~III and Psc~II with respect to one another in the MW-only (dashed lines) and MW-LMC potential (solid lines). In the MW-only potential, the satellites reach as close 20~kpc to each other in the last 2--3~Gyr. For the heavy MW-only potential, Peg~III and Psc~II remain significantly far from each other until $\sim$2~Gyr ago, when they make a close ($<$20~kpc), pericentric passage about both the MW and then subsequently each other. However, for the light MW-only potential, the UFDs have remained within 80~kpc of each other for at least the last few billion years. 

When the influence of the LMC is included, an opposite trend is observed. For the light MW model, the UFDs first interact with one another about 3~Gyr ago (prior to the infall of the LMC) and then pass about each other again 1~Gyr ago (after the infall of the LMC). They are otherwise significantly separated. In the heavy MW+LMC model, the UFDs are within about 80~kpc of each other for the last 6~Gyr. They pass around each other twice at a distance of $\sim$20~kpc at $\sim$1 and $\sim$2~Gyr ago. These results, summarized in Tables \ref{tab:MWonly} and \ref{tab:MWLMC}, illustrate that Peg~III and Psc~II may have had an intricate orbital history that includes close interactions with the MW, LMC, and each other when the \cite{mcconnachie21b} values are considered. These conclusions are based on the direct orbital histories, yet the results in Table \ref{tab:MWonly} and \ref{tab:MWLMC} illustrate that there is significant statistical uncertainty on the properties of this system due to large measurement uncertainties on distance and proper motion. 


We also calculate a simple metric to test whether Peg~III and Psc~II are currently bound to each other as was done in \cite{Geha15} and \cite{Sohn20} for NGC~147 and NGC~185, two dwarf elliptical galaxies located about 1$^{\circ}$ from each other. It is known that for two point masses to be gravitationally bound, the potential energy of the system must be greater than the kinetic energy. This yields the criterion $b \equiv 2 G M_{sys}/\Delta r \Delta v^2$. Thus, when $ b >1$, the system is considered bound. 

Assuming a relative Galactocentric distance of $\Delta r {=} 36.4$ kpc between the dwarfs, a relative Galactocentric velocity of $\Delta v{=}20.6$~km~s$^{-1}$ at present day, and a mass of  $10^9$~M$_{\odot}$ for each dwarf, we find $b{=}1.11\pm0.31$. This $b$ corresponds to the initial conditions used to calculate the direct orbital histories and the uncertainty represents the standard deviation across $b$ calculated for 1,000 initial conditions encompassing the measured phase space of these two dwarf galaxies. Thus, Peg~III and Psc~II are consistent with being bound to each other, as illustrated by the orbital histories in the bottom panels of Figure \ref{fig:orbits}. However, the uncertainty on $b$ indicates that some unbound orbits are possible within the measured phase space. This, in combination with the non-zero transverse motions of these dwarfs, indicates that improved distances and PMs, in addition to a more precise understanding of the MW's mass, are therefore necessary to definitively conclude whether Peg~III and Psc~II have experienced a shared orbital history over the last few billion years.

If Peg~III and Psc~II are indeed bound to each other, as suggested by the most common orbits recovered from the \citet{mcconnachie21b} PMs, this would add to the rare findings of confirmed pairs of satellites within the Local Group. It has also been shown that pairs and groups of satellites that fall into the halo of their hosts together tend to merge with one another, sometimes only 1--3 Gyr after infall \citep[e.g.,][]{Deason14,Wetzel15}. Peg~III and Psc~II could thus be in the process of or on the way to the early phases of merging. 
Conversely, another possibility would be for the two UFDs to disperse in phase space soon after falling into the MW, depending on how strongly the two were bound during infall \citep[e.g.,][]{Deason2015}.
Improved PMs will continue to shed light on the potential future trajectories of these UFDs.

\subsection{Need for Improved PM Measurements}
Based on \textit{Gaia} eDR3 measurements alone, it is unlikely for the possibility of an association between Peg~III and Psc~II to be further constrained.
\textit{Gaia} will operate much longer, however, optimistically achieving a baseline of 11 years. Since proper motion errors of continuously observing telescopes scale with time to the power of $-1.5$ (e.g., \citealt{Lindegren_21a}), the precision achievable with the final \textit{Gaia} data release would be approximately 36 and 30~km~s$^{-1}$ for Peg~III and Psc~II, respectively. 
This precision could already be met and surpassed using \textit{HST} now. We estimate the achievable precision using the data described here as a first epoch and with a second epoch taken as outlined in \citet{Kallivayalil_15}.
A new measurement with \textit{HST} imaging taken in 2023 (a six-year baseline) would measure the transverse velocity to a precision in both dimensions of 30.3~km~s$^{-1}$ for Peg~III and 24.5~km~s$^{-1}$ for Psc~II. If a velocity difference less than $\sim$50~km~s$^{-1}$ was observed, we would be able to exclude the possibility of a chance association at a confidence level of at least 95\% (see Appendix \ref{app:prec} for detailed calculations).

\section{Conclusions}\label{sec:conc}
Based on deep \textit{HST} imaging and \textit{Gaia} eDR3 PMs, we have measured the structural parameters and performed an orbital analyses for two distant MW UFD satellites, Peg~III and Psc~II.
For Peg~III, we measured an elliptical half-light radius of $1.88^{+0.42}_{-0.33}$ arcminutes, a position angle of $85\pm8$ degrees, and an ellipticity of $0.36^{+0.09}_{-0.10}$. 
The best-fit exponential model for Psc~II gave an elliptical half-light radius of $1.31^{+0.10}_{-0.09}$ arcminutes, a position angle of $97\pm3$ degrees, and an ellipticity of $0.37\pm0.04$. Our Psc~II measurements are within $1\sigma$ agreement when compared to previous literature values, while we find a larger elliptical half-light radius value for Peg~III. See Tables \ref{tab:peg3} and \ref{tab:psc2} for full comparisons. Future imaging of similar depth to ours paired with a larger FOV would help clarify the true size of Peg~III.

We compared the measured sizes and magnitudes of Peg~III and Psc~II to both those of other observed faint MW satellites and simulations of isolated field UFDs and satellite UFDs in a MW-like environment. While Peg~III and Psc~II are more distant MW UFDs with $r_h$ and $M_V$ values that correspond well to simulations of isolated field UFDs, they are not significantly distinct in structure from other observed UFDs in the inner MW halo. This could point to Peg~III and Psc~II (and other outer-halo UFDs) having been subject to tidal forces from the MW and/or the LMC throughout their lifetimes.

The first orbital analysis, using solely the \cite{mcconnachie21b} PM values, demonstrated the importance of including the LMC in such calculations, as it affects the timing and pericenter of the orbits of both UFDs. In this analysis, a statistically significant fraction of the computed orbital histories in the measured phase space show Peg~III and Psc~II are consistent with being a gravitationally bound pair today. 

Comparing the vastly different results from the \cite{Li21} study versus our analysis using the \cite{mcconnachie21b} Psc~II PM values and considering that the Peg~III PM measurement came from a single star, it is also clear that improved measurements based on more member stars and a clearer knowledge of the MW's mass are necessary to narrow down the possible orbital histories and allow a more definitive statement on whether Peg~III and Psc~II are indeed a bound pair.
These improved measurements could be taken from imaging by \textit{HST} now or measured from the final \textit{Gaia} data release in years to come. With the precision achievable from these baselines, measuring a velocity difference between Peg~III and Psc~II of less than $\sim$50 km s$^{-1}$ would suggest with about 95\% confidence that these UFDs are bound. This would finally resolve the question and confirm if these two fascinating galaxies are indeed bound, thereby providing a critical data point for understanding dwarf galaxy evolution.

\section*{Acknowledgements}
\begin{acknowledgments}
We would like to thank the anonymous referee for their thorough reading of this manuscript and comments that have led to its improvement and clarification.
These data are associated with the \textit{HST} Treasury Program 14734 (PI: Kallivayalil). Support for this program was provided by NASA through grants from the Space Telescope Science Institute. This material is based upon work supported by the National Science Foundation under grant No. AST-1847909. 
HR acknowledges support from the Virginia Space Grant Consortium Graduate Research STEM Fellowship.
EP acknowledges support from the Miller Institute for Basic Research in Science at University of California, Berkeley.
This research has made use of NASA’s Astrophysics Data System.
\end{acknowledgments}
\textit{Facility}: {HST (ACS,WFC3)}

\textit{Software}: 
Aplpy \citep[][]{aplpy};
Astrodrizzle \citep{Fructer2002};
Astropy \citep{astropy:2013, astropy:2018};
corner.py \citep{corner};
dustmaps \citep{Green2018};
emcee \citep{ForemanMackey2013};
Jupyter Notebook \citep{soton403913};
Matplotlib \citep{Hunter:2007};
Numpy \citep{harris2020array};
photutils \citep{Bradley2020};
scikit-learn \citep{scikit-learn}
Scipy \citep{2020SciPy-NMeth};
stsynphot \citep{stsynphot};
synphot \citep{synphot}

\appendix
\section{Calculations of Current and Future Precision} \label{app:prec}

To formally investigate the precision needed for determining an association, we perform a Monte Carlo simulation, drawing the transverse velocities in both dimensions (X and Y) for two galaxies in both an unassociated and associated case. To do this, we must consider the dispersion of the halo ($\sim$100~km~s$^{-1}$; \citealt{Correa-Magnus_21}) as well as the proper motion uncertainties. \textit{Gaia's} current measurement errors for the two UFDs are clearly larger than the halo dispersion: without using a prior (which artificially constrains the error), the best measurement determined for Psc~II using \textit{Gaia} eDR3 has a precision of about 230~km~s$^{-1}$ \citep{Li21}, while a proper motion uncertainty cannot even be obtained for Peg~III. Nonetheless, to calculate an estimate of the necessary precision for determining the presence of an association, we optimistically scale the \cite{Li21} proper motion error by the \cite{Kim2016} distance to obtain 270~km~s$^{-1}$ for Peg~III. 

In the unassociated case, transverse velocities for two galaxies are drawn from a Gaussian distribution centered on 0~km~s$^{-1}$, with $1\sigma$ equal to the halo dispersion of 100~km~s$^{-1}$ plus the measurement errors of each galaxy, as described in the previous paragraph. 

In the associated case, we again center the Gaussian distribution on 0~km~s$^{-1}$, but instead set $1\sigma$ equal to only the current \textit{Gaia} eDR3 measurement errors of Peg~III and Psc~II (i.e., without including the halo dispersion). We then calculate the velocity difference for each realization:
\begin{equation}
    V_{\mathrm{diff}}{=}\sqrt{(V_\mathrm{Psc\,II,X}-V_\mathrm{Peg\,III,X})^2+(V_\mathrm{Psc\,II,Y}-V_\mathrm{Peg\,III,Y})^2}.
\end{equation}

We take the median $V_\mathrm{diff}$ from the associated case and compare it to the $V_\mathrm{diff}$ values calculated in the unassociated case and determine how often the unassociated case produces a velocity difference less than or equal to the median associated $V_\mathrm{diff}$. We obtain a median $V_\mathrm{diff}${=}418~km~s$^{-1}$ from the associated realizations. A velocity difference of this size or smaller occurs with a 45\% probability in the unassociated case. Thus, it is not expected that \textit{Gaia} eDR3 measurements without a prior could give any constraints on whether or not Peg~III and Psc~II are associated.

Now we investigate how future measurements could improve upon current information to constrain an association between Peg~III and Psc~II. 

Repeating the Monte Carlo simulation with the estimated \textit{HST} errors from a six-year baseline (Peg~III: 30.3~km~s$^{-1}$; Psc~II: 24.5~km~s$^{-1}$), we obtain a median $V_\mathrm{diff}${=}46~km~s$^{-1}$ from the associated cases. A velocity difference of this size or smaller occurs with a 4.7\% probability in the unassociated case. 
Thus, for an observed velocity difference of less than $\sim$50~km~s$^{-1}$, the small probability from the unassociated Monte Carlo cases, combined with the already small probability of similar line-of-sight velocities and distances (2.5\%, see Section~\ref{subsec:bound}), would almost completely rule out a chance association.

Given a longer baseline, \textit{HST}(-like) observations taken with
\textit{HST}, \textit{JWST}, or the \textit{Roman Space Telescope} could measure the velocity well enough to prove that the two systems are associated with each other. 
For example, observations taken in 2034 would give a full transverse velocity error of 19.4~km~s$^{-1}$, smaller than the current Galactocentric velocity difference ($\sim$20.6~km~s$^{-1}$).

\bibliography{peg3psc2.bib}
\bibliographystyle{aasjournal}

\end{document}